\documentclass[aps,prd,preprint,groupedaddress,showpacs,nofootinbib]{revtex4}

\usepackage{amsmath}
\usepackage{amssymb}
\usepackage{mathtools}
\usepackage{amsthm}

\theoremstyle{plain}

\usepackage{graphicx}
\usepackage[justification=justified]{subcaption}
\usepackage{caption}
\usepackage[colorinlistoftodos]{todonotes}

\usepackage{hyperref}
\usepackage{bbold}
\usepackage{cancel}
\usepackage{xcolor}

\usepackage{enumerate}

\DeclareGraphicsExtensions{.pdf,.png,.jpg,.eps}
\usepackage[utf8]{inputenc}
\usepackage{amsmath}
\usepackage{amsfonts}
\usepackage{amssymb}
\usepackage{txfonts}
\usepackage{bbold}
\graphicspath{{TeXFigures/}}

\begin{document}
\unitlength = 1mm

\title{Inflation as a spontaneous symmetry breaking of Weyl symmetry}

\author{Alexander Barnaveli}
\email{a.barnaveli@students.uu.nl}

\author{Stefano Lucat}
\email{s.lucat@uu.nl}

\author{Tomislav~Prokopec}
\email{t.prokopec@uu.nl}

\affiliation{Institute for Theoretical Physics, Spinoza Institute and EMME$\Phi$, Utrecht University,\\
Postbus 80.195, 3508 TD Utrecht, The Netherlands}

\date{\today}

\begin{abstract}
In this paper we study a novel realization of inflation, based on Weyl invariant 
gravity with torsion. We show that requiring the classical action for the scalar field
to be Weyl invariant introduces a dilaton which induces a non trivial modification of the field space geometry of 
the scalar sector, which allows for inflationary phase that 
begins at the conformal point of the inflaton $\psi$, {\it i.e.} $\langle\psi\rangle=0$. Since the model is 
Weyl invariant, the inflaton condensation models a process of spontaneous Weyl symmetry breaking. 
For a wide range of parameters the spectral observables of the model are in good agreement with the CMB measurements, such
that the scalar spectral index and the tensor-to-scalar ratio approximately agree with those of Starobinsky's inflation, {\it i.e.}
$n_s\simeq 0.96-0.97$ and $r \approx 3\times 10^{-3}$. The simplest version of our model contains two scalar degrees of freedom, one of them being an exactly flat direction. If that degree is excited early on in inflation and if inflation 
lasts for about 60 e-foldings,
we find that the Unverse undergoes a short period of kination that predates inflation. Such a period strongly suppresses the amplitude of 
large scale CMB temperature fluctuations providing thus an elegant explanation for the lack of power in the lowest CMB multipoles.

\end{abstract}


\maketitle

%%%%%%%%%%%%%%%%%%%%%%%%%%%%%%%%%%%%%%%%%%%%%%%%%%%%%%%%%%%%%%%%%%%%%%%%%
%%%%%%%%%%%%%%%%   I N T R O D U C T I O N   %%%%%%%%%%%%%%%%%%%%%%%%%%%%
%%%%%%%%%%%%%%%%%%%%%%%%%%%%%%%%%%%%%%%%%%%%%%%%%%%%%%%%%%%%%%%%%%%%%%%%%

\section{Introduction}
\label{introduction}

The phenomena known as cosmic inflation~\cite{Starobinsky:1980te,Guth:1980zm} is one of the 
most studied in high energy in the modern days~\cite{Martin:2013tda}. 
The leading paradigm to study it is to construct a 
so-called effective field theory based on the breaking of time 
translation symmetry induced by the expansion of the universe~\cite{Cheung:2007st}.

While this point of view provides a self consistent phenomenological description,
it does not shed any light on the physical nature of the phenomena, nor it reveals 
any of the basic principles that rule its dynamics. A possibility to gain understanding
is to realize cosmic inflation in theories with enhanced symmetry, which are
broken today. This would then lead to a phase transition of 
sort, in which the rolling inflaton field acts as the order parameter in the 
symmetry breaking process.

 A popular choice in this direction is to invoke scale or conformal symmetry
 as the one realized in the initial state of the universe, 
 and broken during the inflationary epoch~\cite{Wetterich:1987fm,GarciaBellido:2011de,Bezrukov:2014ipa,Karananas:2016grc,Kallosh:2013maa,Kallosh:2013hoa,Ozkan:2015kma,Carrasco:2015rva,Ferreira:2016vsc,Ferreira:2016wem,Ferreira:2018qss,Rubio:2017gty,Kannike:2016wuy,Salvio:2017qkx,Salvio:2017xul,Salvio:2014soa,Rinaldi:2015uvu,Karananas:2016kyt,Casas:2017wjh,Tang:2018mhn,Achucarro:2017ing,Klein:2017hch}. This is also the approach advocated in this paper, however with an important twist. 
 
 The reasons why one expects conformal symmetry to be restored at very early times
 of the cosmic history are multiple. A compelling argument is given by the idea that 
 quantum field theories at very short distances become approximately conformal and 
 flow towards conformally invariant theories, the so-called fixed points of the renormalization 
 group flow. Such a quantum theory can be, at least in principle, rigorously defined, 
which is what makes it so appealing. 
Another theoretical argument is provided by the simplicity that 
 scale invariant models enjoy. Namely, the symmetry permits a handful 
 of (local) operators that can be included, such that these theories differ from one another only 
 by the number of degrees of freedom they possess and by their spin. 
 
In this paper we consider a generic choice for the scale invariant 
theory, containing $\mathcal{N}$ scalar degrees of freedom~($\in O(\mathcal N)$), and in which the
global scale symmetry is promoted to a local~(gauged Weyl) invariance. Our conclusion are general
in the sense that they do not depend on the specific of the theory, 
but only upon considering a Weyl invariant theory.
In order to obtain a gauge invariant action, a compensating Weyl one form is 
introduced in the theory, which we interpret as the torsion tensor 
trace~(as was argued in~\cite{Lucat:2016eze}). 

The longitudinal component of the compensating one form acts as the Goldstone
mode of the broken symmetry, and effectively behaves as a scalar field that kinetically 
mixes with the other scalars in the theory. The Weyl invariance of the underlying theory 
enlarges the configuration space of scalar fields from the original $O(\mathcal N)$ invariant scalars 
to a negatively curved, $\mathcal N +1$ dimensional 
field space, the hyperbolic space $\mathbb{H}^{\,\mathcal{N}+1}$. This is similar to 
the hyperbolic field geometry studied in~\cite{Kallosh:2013maa},
where the authors consider it motivated by supersymmetry. That has also been discussed 
in~\cite{Karananas:2016kyt}, where the authors prove 
that a maximally symmetric field space leads to universal predictions, which might be responsible for the universality properties 
discussed in~\cite{Kallosh:2013hoa}, thus justifying the name $\alpha$-attractors. 
The physical consequence of the negative curvature of the configuration space is 
to stretch the potential at the boundary of the field space allowing for a plateau on which a slow roll
inflation is possible. In the realization of~\cite{Kallosh:2013maa} the potential is stretched at large 
field values, $\phi\rightarrow \infty$, while in our realization the stretching occurs near the origin of the inflaton direction, {\it i.e.} $\phi\sim 0$. {Another interesting 
effect a negative configuration space curvature 
can have was discussed in Ref.~\cite{Renaux-Petel:2015mga}, 
where it was pointed out that a negative curvature can reduce the effective mass of fields 
during inflation, making them even negative, and thus qualitatively change the inflationary dynamics.
This mechanism was dubbed geometric destabilization.  
}

The additional scalar direction given by the Goldstone mode is flat in the sense that the Goldstone mode is only derivatively coupled to the
$O(\mathcal N)$ scalars, as we would expect from a Goldstone mode. 
If its energy density is initially big, it can dominate the preinflationary epoch 
with possible observable consequences. However, it does not play a dominant role 
in the inflationary dynamics and can thus be neglected at any later time. 

Finally we briefly study quantum corrections from the matter sector, by computing the one loop quantum effective action,
and conclude that the hierarchy required to obtain inflationary observables compatible 
with the most recent cosmological data~\cite{Aghanim:2018eyx}
is stable. For the same reason a late time small dark energy can be expected
and maintained small, thus making the tiny observed cosmological constant 
in this model technically natural~\cite{Lucat:2018slu}.

The paper is organized as follows: we review the Weyl invariant theory of 
gravity in section~\ref{geometric.theory.review} and then show how the negatively curved field
space geometry is induced from the requirement that the theory is Weyl invariant. 
In section~\ref{Inflationary dynamics} we study the inflationary dynamics and 
explain the dynamics of the flat direction $\chi$, which constitutes the Goldstone
mode of the broken symmetry. In section~\ref{Results} we discuss the model 
predictions, and discuss how we could detect the Goldstone mode $\chi$ in
the CMB spectrum. Finally in section~\ref{Quantum.Corrections} we briefly 
discuss the quantum corrections to the model. 
In the Appendix one can find a comparison between this model and the Abelian 
symmetry breaking phenomenon, where we highlight the differences among the two.

\section{Weyl invariant gravity with torsion}
\label{geometric.theory.review}

Defining the torsion tensor as the antisymmetric part of the connection,
\begin{equation}
\label{torsion.definition} T^\lambda{}_{\mu\nu}  = \Gamma^\lambda{}_{[\mu\nu]}\,,
\end{equation}
one finds an exact gauge symmetry of curvature and the geodesic equation,
if the metric and the torsion are transformed according to~\cite{Lucat:2016eze},
\begin{equation}
\label{transformation.law.Weyl} T^\lambda{}_{\mu\nu} \rightarrow T^\lambda{}_{\mu\nu} + \delta^\lambda_{[\mu}\partial_{\nu]} \theta\,,\quad
g_{\mu\nu}\rightarrow e^{2\theta}g_{\mu\nu}\,.
\end{equation}
That is, defining the torsion trace as,
\begin{equation}
{\cal T}_\mu \equiv -\frac{2}{D-1} T^\lambda{}_{\mu\lambda}
\,,
\label{torsion trace}
\end{equation}
one finds it transforms locally under~(\ref{transformation.law.Weyl}) as a $U(1)$ gauge field, 
\begin{equation}
{\cal T}\rightarrow {\cal T} + \text{d}\theta
.
\label{transformation of torsion trace}
\end{equation}
The transformation~(\ref{transformation.law.Weyl}) also realises a reparametrization of proper time,
$\text{d}\tau \rightarrow e^{\theta} \text{d}\tau$ which leaves the geodesic equation invariant~\cite{Lucat:2016eze}.
Globally, however, the group of Weyl transformations is inequivalent to that of an Abelian gauge  group 
as the corresponding group space can be 1-to-1 mapped onto the set of real numbers and it is thus non-compact.

In Ref.~\cite{Lucat:2016eze} we also showed how to straightforwardly extend such a symmetry to the classical
standard model Lagrangian, which modifies scalar kinetic terms 
according to the prescription~\footnote{Here and thereafter we use the notation $\nabla_\mu$ to 
denote the covariant derivative which commutes both with diffeomorphisms and Weyl
transformations~(\ref{transformation.law.Weyl}). Its definition when acting on tensors can be found 
in~\cite{Lucat:2016eze}, and it agrees with~(\ref{covariantization.prescription}) when acts on a scalar},
\begin{equation}
\label{covariantization.prescription} 
\partial_\mu \phi \rightarrow (\partial_\mu - \Delta_\phi {\cal T}_\mu)\phi\equiv \nabla_\mu \phi\,,
\end{equation}
where the $\Delta_\phi$ is the scaling dimension of the field $\phi$. For canonically normalized scalars,
it evaluates to $\Delta_\phi =-(D\!-\!2)/2 $. 

With this prescription and using the fact that the Ricci scalar with torsion transforms under~(\ref{transformation.law.Weyl}) 
as, $R\rightarrow e^{-2\theta}R$, the Weyl invariant operators that constitute the action in four dimensions are limited to,
\begin{eqnarray}
\label{Weyl.invariant.action}
 S \!&=&\! 
\int \text{d}^4x\sqrt{-g} \Bigg [\alpha R^2 
+ \zeta R_{(\mu\nu)} R^{\mu\nu} + \frac{\xi}{2}  \phi^I \phi^J \delta_{IJ}R
 - \frac\lambda4 \left ( \phi^I \phi^J \delta_{IJ}\right )^2 \\
 \!&-&\! 
\frac{1}{2}g^{\mu\nu}\delta_{IJ} \left (\partial_\mu\phi^I 
- \Delta_\phi {\cal T}_\mu\phi^I\right) \left (\partial_\nu\phi^J { -} \Delta_\phi {\cal T}_\nu\phi^J\right) 
- \frac{\sigma}{4} \mathcal{T}_{\mu\nu}\mathcal{T}^{\mu\nu}\Bigg ]\,,\nonumber\\
 &&
\mathcal{T}_{\mu\nu} \equiv \partial_\mu {\cal T}_\nu - \partial_\nu {\cal T}_\mu\,,
\end{eqnarray}
where $\alpha,\zeta,\xi,\lambda$ and $\sigma$ are dimensionless couplings 
and we allow for $O(\mathcal{N})$ invariant scalars $\phi^I$, with $I = 1,\cdots\,,\mathcal{N}$.~\footnote{ 
In principle, we could allow for a more generic field space metric, $\mathcal G{}_{IJ}$
with a different symmetry group. Simple considerations on scale invariance show, however, that 
such a metric can only depend on ratios $\phi^A/\phi^B$, thus lowering the symmetry 
to a subgroup of $O(\mathcal{N})$. Since in this paper we follow a logic of simplicity, we are going to
choose the maximally symmetric option, that is $O(\mathcal{N})$ for which $\mathcal{G}_{IJ}=\delta_{IJ}$.
Quantum corrections will in general modify the kinetic term such to replace $\delta_{IJ}$ by a more general 
 $\mathcal{G}_{IJ}$ of the form $\mathcal{G}(\phi^K \phi^L \delta_{KL}/\mu^2)\delta_{IJ}$, where $\mu$ is a 
renormalization scale,  
which still respect the $O(\mathcal{N})$ symmetry but mildly breaks 
the Weyl symmetry. Furthermore, one could imagine
$\mathcal G{}_{IJ} = \eta_{IJ}$, having one or more time-like directions. This would however introduce ghost-like directions
in field space, which would destabilize the field dynamics, and for that reason we shall not allow that possibility.
} 

The study of the action~(\ref{Weyl.invariant.action}) is best carried out in the Einstein frame,
that is where gravity follows the Einsteinian dynamics.
This can be achieved by setting the parameters $\zeta$ and the torsion field strength coupling $\sigma$
to {\it zero}. 
While the former is necessary as it renders the model classically stable, the latter is not required, but it is convenient as it removes the dynamical modes of the torsion trace vector, thus simplifying the model.
With this choice the transverse components of the torsion trace become constraint fields which can be consistently set to zero. 
The remaining longitudinal component of ${\cal T}_\mu$ can be modeled by a real scalar field $\phi^0$ as follows, 
\begin{equation}
{\cal T}_\mu=\partial_\mu\phi^0
\,.
\label{phi 0}
\end{equation}
This choice is discussed in detail in the Appendix, in which we discuss at length the Weyl 
gauge fixing and explain the differences between the local Weyl symmetry 
and the local Abelian gauge symmetry. 

The Einstein's frame action, which is on-shell equivalent to~(\ref{Weyl.invariant.action}), reads,
\begin{equation}
\begin{split}
\label{einstein.frame.action}
 S_E = \int \!\text{d}^4x \sqrt{-g} \Bigg [&\!\!-\!\Big(\frac{\xi^2}{16\alpha}\!+\! \lambda\Big)\left (\delta_{IJ}\phi^I \phi^J\right )^2 
+\frac{\xi}{8\alpha}\omega^2\delta_{IJ}\phi^I \phi^J 
+\frac{\omega^2}{2}\Big(\overset{\circ}{R}\!+\!6\overset{\circ}{\nabla}{}^2\phi^0
                            \!-\! 6\partial_\mu\phi^0\partial^\mu\phi^0\Big)
\\
-\frac{\omega^4}{16\alpha}
&
- \frac{1}{2} \delta_{IJ}g^{\mu\nu} \left(\partial_\mu\!+\!\partial_\mu\phi^0\right) \phi^I \left(\partial_\nu\!+\!\partial_\nu\phi^0\right) \phi^J
\Bigg ]\,,
%I\,,J= 1,\cdots,\mathcal{N}\,,
\\
\equiv \int \text{d}^4x \sqrt{-g} \Bigg [&\!-\!\Big(\frac{\xi^2}{16\alpha}\!+\! \lambda\Big)
\left (\delta_{IJ}\phi^I \phi^J\right )^2 +\frac{\xi}{8\alpha}\omega^2\delta_{IJ}\phi^I \phi^J
+ \frac{\omega^2}{2}\overset{\circ}{R}
 -\frac{\omega^4}{16\alpha}              
+3\omega^2\overset{\circ}{\nabla}{}^2\phi^0
\\
 &
- \frac{1}{2} \mathcal{G}_{AB}g^{\mu\nu} \partial_\mu\phi^A \partial_\nu\phi^B 
\Bigg ]
\,,\;
A\,,B= 0,1,\cdots,\mathcal{N}\,,
 \end{split}
 \end{equation}
where  $\omega$ and $\theta$ are Lagrange multiplier fields,
 $\mathcal{G}_{AB}$ is an extended configuration space metric which includes the longitudinal torsion direction $\phi^0$,
\begin{equation}
{\cal G}_{00}=6\omega^2+\delta_{IJ}\phi^I\phi^J
\,,\qquad {\cal G}_{0I}=\delta_{IJ}\phi^J={\cal G}_{I0}
\,,\qquad
{\cal G}_{IJ}=\delta_{IJ}
\,,\quad (I,J=1,\cdots \mathcal{N})
\,,
\nonumber
\end{equation}
and we have also substituted the expression for the Ricci scalar with torsion,
\begin{eqnarray}
R^\lambda{}_{\alpha\sigma\beta} \!&=&\! \partial_\sigma \Gamma^\lambda{}_{\alpha\beta} - \partial_\beta  \Gamma^\lambda{}_{\alpha\sigma} +   \Gamma^\lambda{}_{\rho\sigma}  \Gamma^\rho{}_{\alpha\beta} -  \Gamma^\lambda{}_{\rho\beta}  \Gamma^\rho{}_{\alpha\sigma} \,,\\
R \!&=&\! g^{\alpha\beta} R^\lambda{}_{\alpha\lambda\beta}
 = \overset{\circ}{R} +6\overset{\circ}{\nabla}{}^{\mu} {\cal T}_\mu 
 - 6 g^{\mu\nu}{\cal T}_\mu {\cal T}_\nu\,.
\end{eqnarray}
Symbols with a $\circ$ on top in~(\ref{einstein.frame.action}) are computed using the metric tensor only, {\it e.g.}
\[\overset{\circ}{\Gamma}{}^\lambda_{\mu\nu} = \frac{g^{\lambda\sigma}}{2}\left (g_{\sigma\mu,\nu}+g_{\sigma\nu,\mu}-g_{\mu\nu,\sigma}\right)\,\]
denotes the Levi-Civit\`a connection.

Varying the action~(\ref{einstein.frame.action}) with respect to $\omega$ yields,
\begin{equation}
\label{constraint.1} 
\omega^2\equiv 4\alpha R+ \xi \phi^2\,,\qquad 
\phi^2\equiv \delta_{IJ}\phi^I\phi^J
\end{equation}
which means that everywhere  in~(\ref{einstein.frame.action}) one can exact the replacement: 
\begin{equation}
\label{constraint.2} 
\omega^2 \;\longrightarrow\; 4\alpha R+ \xi \phi^2\,,
\end{equation}
which will give back the original Jordan frame action~(\ref{Weyl.invariant.action}).  

The action~(\ref{einstein.frame.action}) still contains a Weyl gauge redundancy. A convenient 
gauge choice is the Weyl gauge which amounts to fixing $\omega$ to some (constant) physical scale. 
From~(\ref{einstein.frame.action}) we see that $\omega$ defines the Planck scale, {\it i.e.} 
\begin{equation}
\label{gauge.fixing} 
\omega^2\equiv 4\alpha R+ \xi \phi^2\;\longrightarrow M_{\rm P}^2\,,
\end{equation}
completely fixing the Weyl gauge. Since now we have defined a reference scale, any other scale in the model
can be defined with respect to that reference scale. This accords with the general notion that 
all physical quantities (or, equivalently, measurements) can be represented as dimensionless ratios.

Note that, because the gauge fixing condition~(\ref{gauge.fixing})
can be imposed only if either $\phi$ or $R$ (or both) does not vanish, we have to take 
the initial conditions such that at least one condensate does not vanish. In other words, 
the conformal point at which all scalar condensates vanish is singular, and a proper discussion of 
its significance is beyond the scope of this work.
In the construction of the mechanism analyzed in this paper, we assume that 
the scalar~(inflaton) field begins at its conformal point, $\langle\phi\rangle = 0$. This choice is motivated by the conformal 
symmetry of the UV theory. However, the initial value for the Ricci scalar is 
such to satisfy~(\ref{gauge.fixing}), which necessarily leads to $R\neq 0$. 
The dynamics that follow are governed by 
a transfer of energy (and entropy) between the space-time fluctuations and the scalar field, as it can be best understood
by considering the scalar potential and noticing that the potential energy of the scalar 
field drops from its value at the beginning of inflation to a lower value when the field reaches the 
global minimum. In this sense, the symmetry breaking described in this paper is due to the initial 
conditions, which are taken at the point of enhanced symmetry. 
Then inflation happens as a consequence of energy exchange between 
the gravitational field and the scalar field. While the Ricci curvature 
breaks conformal invariance even near $\phi = 0$, it is still conceivable that 
conformal symmertry is realized in the far UV, {\it i.e. } at scales much greater than $\omega=M_{\rm P}$.
A proper understanding of the UV conformal fixed point is, however, beyond the scope of this 
work.~\footnote{Curiously, if initial conditions support cosmic inflation, then 
its geometry can be approximated by that of de Sitter space, which exhibits a global conformal symmetry SO(2,4).
}

Coming back to our inflationary model, the extended {\it configuration space metric} in~(\ref{einstein.frame.action}) reads,
\begin{equation}
{\cal G}_{AB}\text{d}\phi^A\text{d}\phi^B = \left (6M_P^2+ \phi^I\phi^J\delta_{IJ} \right ) (\text{d}\phi^0)^2+ 2\phi_I\text{d}\phi^I\text{d}\phi^0 +\delta_{IJ}\text{d}\phi^I\text{d}\phi^J\,,
\end{equation}
and has a negative configuration space Ricci curvature, ${\cal R}=-\frac{\mathcal{N}(\mathcal{N}+1)}{6 M_P^2}$, which identifies it 
as a hyperbolic geometry $\mathbb{H}^{\,\mathcal{N}+1}$. 

To see this more explicitly, consider the following coordinate transformations
in field space. First, let us define polar coordinates for the $O(\mathcal{N})$ scalars,
\begin{equation}
\begin{split}
\phi^1 =& \rho \cos\theta_1
\,,\quad
\phi^2 = \rho \sin\theta_1\cos\theta_2\,,\quad
\phi^3 = \rho \sin\theta_1\sin\theta_2 \cos\theta_3\,,\;\;\cdots\,,
\\
\phi^{\mathcal{N}-1} =& \rho \sin\theta_1\sin\theta_2\cdots\sin\theta_{\mathcal{N}-2} \cos\theta_{\mathcal{N}-1}
\,,\quad
\phi^{\mathcal{N}} = \rho \sin\theta_1\sin\theta_2\cdots\sin\theta_{\mathcal{N}-2} \sin\theta_{\mathcal{N}-1}
\,,\;\;
\phi^I \phi^I = \rho^2
\\ 
\implies & \phi^I \text{d}\phi^I = \rho\text{d}\rho
\,,\quad 
\label{extended.metric.polar} 
{\cal G}_{AB}\text{d}\phi^A\text{d}\phi^B = 
\left (6M_P^2+ \rho^2 \right ) (\text{d}\phi^0)^2+ 2\rho\text{d}\rho\text{d}\phi^0 +\left( \text{d}\rho^2 + \rho^2 \text{d}\Omega_{\mathcal{N}-1}\right)\,,
\end{split}
\end{equation}
where $\text{d}\Omega_{\mathcal{N}-1}$ is the metric on the $(\mathcal{N}-1)$-sphere, $S^{\mathcal{N}-1}$. 
Finally, defining, 
\begin{equation}
\label{last.change} \phi^0 = \chi - \frac{1}{2} \log\left(1 + \frac{\rho^2}{6M_P^2}\right )\,,
\end{equation}
brings the metric into a diagonal form,
\begin{equation}
\label{manifestly.hyperbolic.metric} 
{\cal G}_{AB}\text{d}\phi^A\text{d}\phi^B 
 = \left (6M_P^2+ \rho^2 \right ) \text{d}\chi^2
             +\frac{6M_P^2\text{d}\rho^2}{6M_P^2 + \rho^2} + \rho^2 \text{d}\Omega_{\mathcal{N}-1}\,,
\end{equation}
which can be further simplified by redefining,
\begin{equation}
\begin{split}
\label{canonical.form.transformation} 
\frac{\text{d}\rho}{\sqrt{6M_P^2 + \rho^2}} &= \text{d}\psi \implies \rho
 =\sqrt{6}M_P \sinh(\psi)\,,\\
\implies &
{\cal G}_{AB}\text{d}\phi^A\text{d}\phi^B 
  = 6M_P^2\left[\text{d}\psi^2+\left (\cosh^2(\psi) \text{d}\chi^2+ \sinh^2(\psi) \text{d}\Omega_{\mathcal{N}-1}\right )\right]
\,.
\end{split}
\end{equation}
This form of the metric makes it explicit what components
of the fields $\phi^I$ are the Goldstone modes and which
are the directions acquiring a {\it vev}, the latter being identified 
with the direction $\psi$ which, as we show in the next section, is 
the inflaton direction. 

\section{Inflationary dynamics}
\label{Inflationary dynamics}

In this section we study inflation in the model presented in the previous section and show that 
one can obtain an inflationary model that conforms with all existing data. 

In the diagonal field coordinates~(\ref{manifestly.hyperbolic.metric}) the gauge fixed action~(\ref{einstein.frame.action})
specialized to $\mathcal{N}=1$ case~\footnote{In this work we focus to study only the simplest O(1) case of a real scalar field.
The more general $O(\mathcal{N})$ case contains $\mathcal{N}-1$ Goldstones, and we postpone the study of their effect onto the inflationary dynamics 
to future publication. It is well known that the dynamics of multi-field Goldstone-like inflationary fields can produce interesting effects, 
see {\it e.g.} Ref.~\cite{Achucarro:2010jv}. 
For now it suffices to note that, when the angular velocities are 
small, $|\dot\theta_I|\ll M_P$ $(I=1,\cdots \mathcal{N}-1)$, we expect the effect of Goldstones to be unimportant during inflation.}, reads,
\begin{equation}
\begin{split}
\label{starting.point.inflationary.dynamics}
S = \int \text{d}^4x \sqrt{-g}\, \Bigg [\!&-\left (\frac{9\xi^2}{4\alpha}
          \!+\!36\lambda\right )M_P^4\sinh^4(\psi) +\frac{3\xi}{4\alpha}M_P^4\sinh^2(\psi) -\frac{M_P^4}{16\alpha}
\\
 &\hskip -0cm 
+ \frac{M_P^2}{2}\overset{\circ}{R}
 - \frac{6M_P^2}{2}  g^{\mu\nu}\left [(\partial_\mu\psi)(\partial_\nu\psi)
                                        + \cosh^2(\psi) (\partial_\mu\chi)(\partial_\nu\chi)\right]
 \Bigg ]
\,.
\end{split}
\end{equation}

Let us begin the analysis by recalling the background cosmological metric,
\begin{equation}
\label{FRW.metric} 
 \text{d}s^2 = - \text{d}t^2 + a^2(t)\text{d}\vec{x}\cdot\text{d}\vec{x}
\,,
\end{equation}
where $a=a(t)$ denotes the scale factor. For simplicity the spatial sections  in~(\ref{FRW.metric}) are assumed to be flat, 
{\it i.e.}
$\text{d}\vec{x}\cdot\text{d}\vec{x} = \delta_{ij} \text{d}x^i\text{d}x^j$ ($i,j=1,2,3$).

\begin{figure}
    \begin{subfigure}[b]{0.46\textwidth}
	\includegraphics[width=\textwidth]{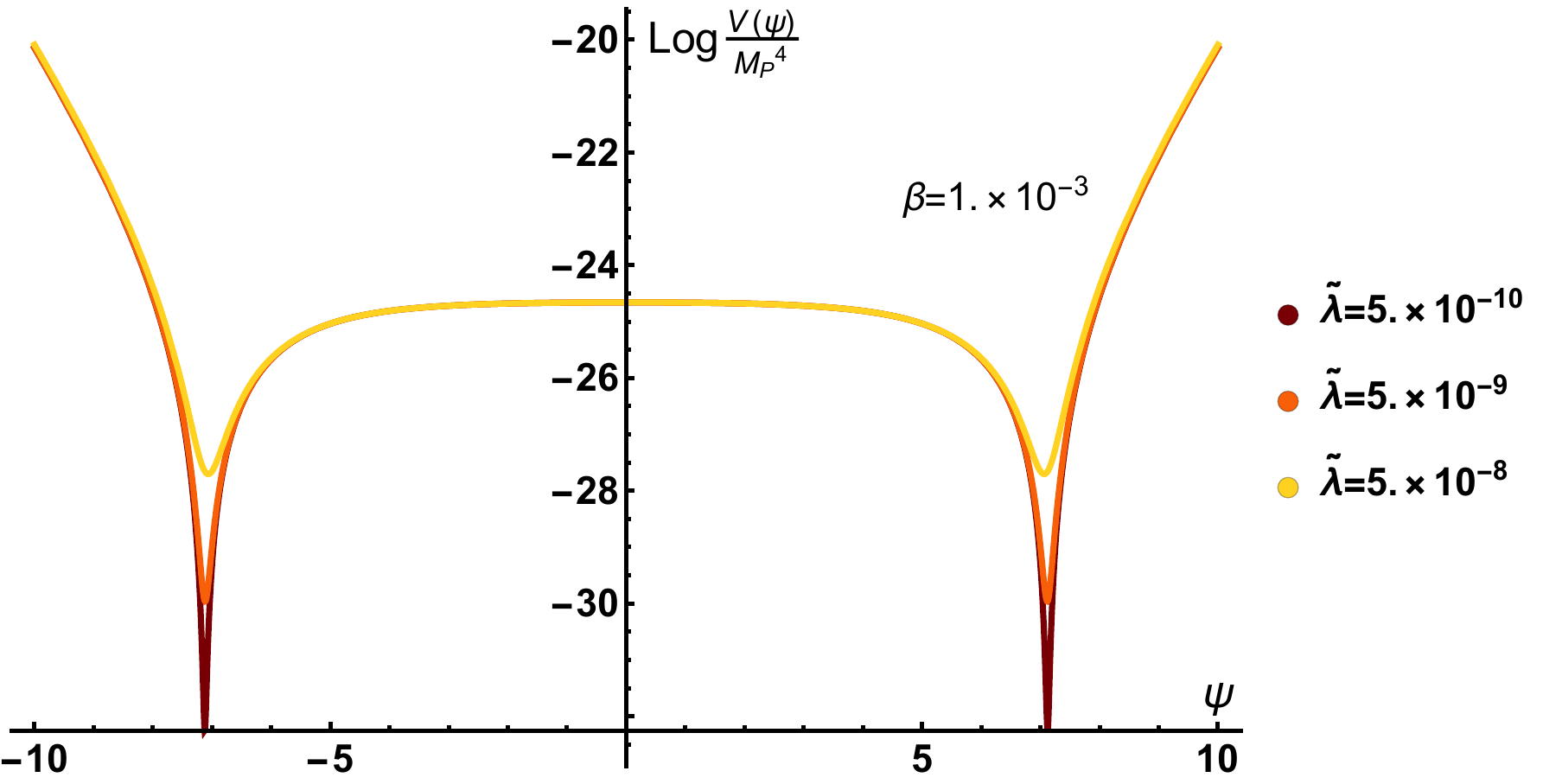}
	\caption{Logarithm of the potential in~(\ref{starting.point.inflationary.dynamics}), for various values of $\lambda$.}
	\label{LogV}
    \end{subfigure}
    \qquad
     \begin{subfigure}[b]{0.46\textwidth}
	\includegraphics[width=\textwidth]{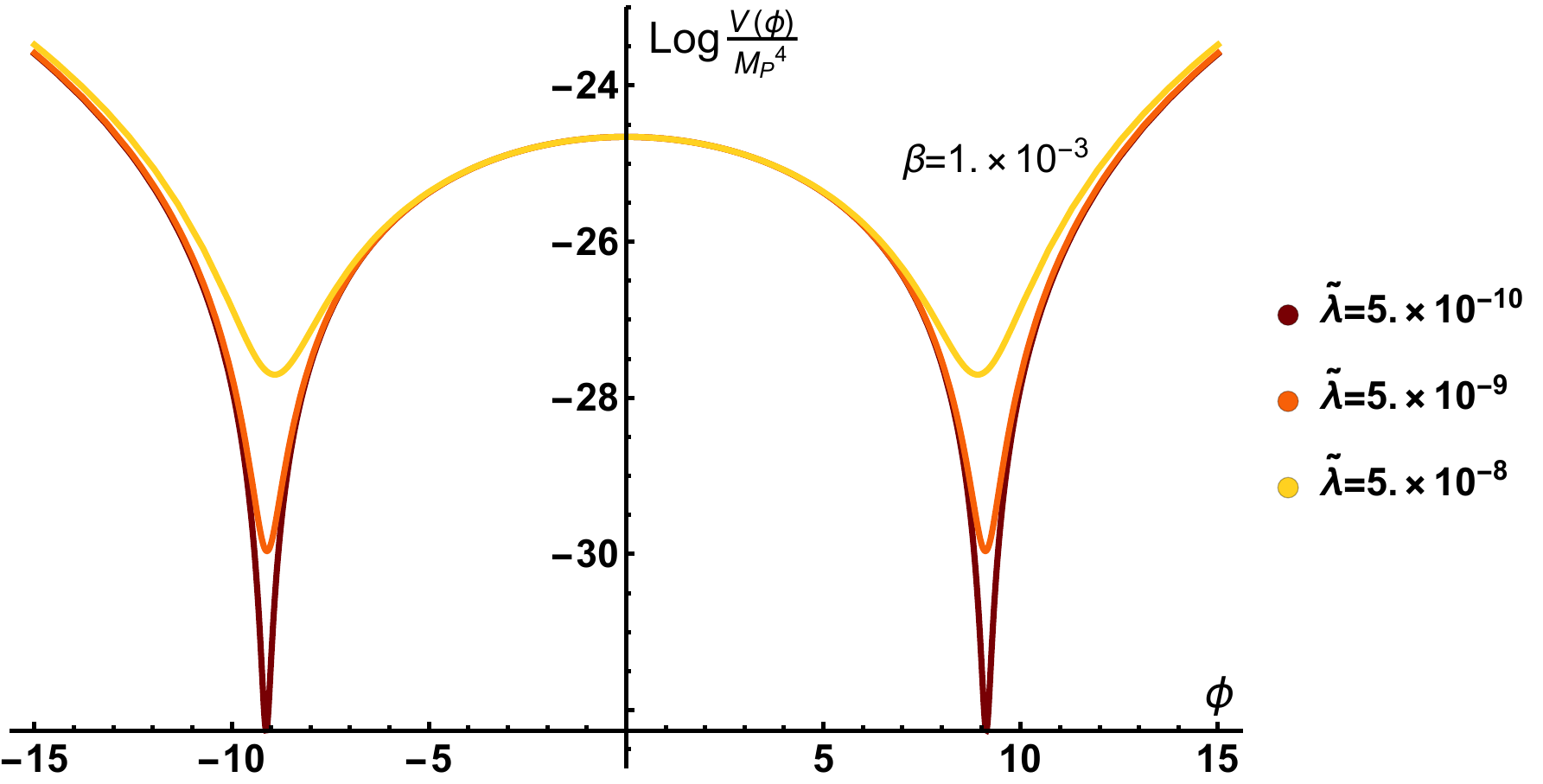}
	\caption{Logarithm of the same potential for a flat field configuration space.}
	\label{LogVflatspace}
    \end{subfigure}
    \captionsetup{justification=centerlast}
\caption{The figure illustrates how the curved geometry of field space allows for a large flat plateau near the origin,
where inflation may take place. The flat region is a consequence of the negatively curved geometry of field space.}
\end{figure}

The equations of motion for the background (homogeneous) fields, $\psi=\psi(t)$ and $\chi=\chi(t)$,
are best analyzed in terms of $e$-folding time, 
$\text{d}N = H \text{d}t$, where $H=\dot a/a$  is the Hubble parameter and $\dot a\equiv da/dt$.
In $e$-folding time the scalar equation of motion and the Friedman equations decouple,
that is defining the principal (first) slow roll parameter as $\epsilon = -\frac{\text{d}}{\text{d}N}\log H\equiv -(\log H)'$,
we find,
\begin{eqnarray}
\label{Friedmann.Equation.efolding.time} 
\epsilon_1\equiv \epsilon\!&=&\!- \frac{H'}{H} = 3\left[(\psi')^2 + \cosh^2(\psi) (\chi')^2\right]
\,,\quad
H^2 =\frac{V(\psi)}{(3\! -\! \epsilon)M_{\rm P}^2}
\,,
\\
\label{Scalar.Equation.efolding.time} \frac{\psi'' }{3-\epsilon} \!&+&\! \psi' 
  +\frac{M_{\rm P}^2}{6} \frac{\partial \log V(\psi)}{\partial \psi}=0
 \,,\quad
\left [H e^{3N} \cosh^2(\psi) \chi'\right ]'=0\,,
\end{eqnarray}
a {\it prime} (${}^\prime$) denotes a derivative with respect to $N$ and 
\begin{equation}
 V(\psi) = M_{\rm P}^4\left[\Big(\frac{9\xi^2}{4\alpha}+36\lambda\Big)\sinh^4(\psi)-\frac{3\xi}{4\alpha}\sinh^2(\psi)
                      +\frac{1}{16\alpha}\right]
\,.
\label{potential for psi}
\end{equation}
The potential~(\ref{potential for psi})
possesses a nearly flat region near the point of enhanced symmetry, $\psi = 0$ (see figure~\ref{LogV}), 
which is instrumental for prolongation of the inflationary phase.
{
An analogous potential that is not however motivated by conformal symmetry is plotted 
on figure~\ref{LogVflatspace} for the same choice of the couplings.
Note that the minimum of the potential in figure~\ref{LogV} is at a lower value of 
the field, implying that the models with a negative configuration space curvature roll typically over smaller distances. 
This feature is due to the sudden curving of the potential that can be seen  in figure~\ref{LogV} 
induced by the negative configuration space curvature. The main consequence of 
such a sharp turn of the potential is, as we shall demonstrate later, a lower value of the slow
roll parameters in the window $N=50-60$ e-foldings before the end of inflation. This 
{translates} into a lower value of the tensor to scalar ratio, $r\simeq 16\epsilon_1$, 
which renders our model viable for a wider range of the parameter $\xi$ as compared
to the flat geometry case. In addition, due to the smaller field excursion in the 
hyperbolic geometry 
{ case~(${\cal O}(1)m_{\rm P}={\cal O}(1)\sqrt{8\pi}M_{\rm P}$),}
 it may be therefore easier to 
tame the Planck scale operators and the infamous $\eta$ problem may be less severe, or even absent, in models 
with a negative configuration space curvature.
}

The potential~(\ref{potential for psi}) has a maximum at $\psi=0$, at which the potential energy equals,
\begin{equation} 
V(0)=\frac{M_{\rm P}^4}{16\alpha}
\,,
\label{V at origin}
\end{equation}
 and two symmetric minima, 
\begin{equation} 
\pm\psi_m=\pm\ln\left[\iota+\sqrt{\iota^2+1}\right]
\,,\qquad
{\rm with} \quad 
\iota=\sqrt{\frac{\xi}{6(\xi^2+16\alpha\lambda)}}
\,, 
\label{psi m}
\end{equation}
at which the potential minimizes at a value,
\begin{equation}
V_m=V(\pm\psi_m)=\frac{\lambda}{\xi^2+16\alpha\lambda}M_{\rm P}^4
\,,
\label{potential minimum}
\end{equation}
such that, as the field rolls down from its maximum at $\psi\approx 0$ to its minimum, the potential energy density changes by,
$\Delta V = -M_{\rm P}^4/[(\lambda/\xi^2)+(16\alpha)^{-1}]$. From~(\ref{potential minimum}) we see that the potential
energy at the end of inflation is positive and potentially large. 
Indeed, unless the coupling constant $\lambda$ is extremely small (and/or $\xi,\alpha$ extremely large),
the energy density~(\ref{potential minimum})  will be much larger than the corresponding density in dark energy. 
Therefore, the amount in~(\ref{potential minimum}) ought to be compensated by an almost equal, negative contribution.
In fact, such a negative contribution exists in the standard model.
Indeed, it was pointed out in Ref.~\cite{Lucat:2018slu} that 
such a fine tuning can be exacted by adding both the negative contributions generated at the electroweak scale 
by the Higgs field condensate and the chiral condensate generated at the quantum chromodynamic (QCD) transition.
In the same reference it was observed that, once this fine tuning is done at one renormalization scale, it will remain stable under 
an arbitrary change of the renormalization scale, which is due to the technical naturalness that arises from 
the enhanced symmetry at the point of vanishing vacuum energy.

A nontrivial consequence of Weyl symmetry, is the absence of any potential for the $\chi$ field.
Since this is a remnant of a broken local symmetry, when quantum
effects are included the $\chi$-flatness must be preserved to all orders in perturbation theory. 
Namely, quantum effects can (and will) modify the term multiplying 
$(\partial\chi)^2$ and they can generate non-local terms,
but they can never generate a potential (zero derivative) term.

An important consequence of the  $\chi$-flatness is the conservation of its canonical momentum,
$\pi_\chi=H e^{3N} \cosh^2(\psi) \chi'$ implied by the equation of motion for 
$\chi$ in~(\ref{Friedmann.Equation.efolding.time}).~\footnote{An analogous conservation of the angular momentum of 
the Goldstone modes governs their dynamics, as each of the Goldstones exhibits a flat direction as well.}
This then means that, 
\begin{equation}
\chi' = \frac{c}{H e^{3N} \cosh^2(\psi) }\,,
\label{c: definition}
\end{equation}
where $c$ is a constant with the meaning of field space angular momentum.
If $c$ is very large, it can change the initial dynamics, producing a period of 
{\it kination} -- defined as the cosmological epoch during which kinetic energy of a scalar field dominates
the Universe's dynamics~\cite{Spokoiny:1993kt,Joyce:1997fc} -- 
during which the matter fluid is characterized  by the equation of state, $P=\rho$.
Since $\chi'$ scales as $1/a^3$ and moreover it is proportional to $1/\cosh^2\psi$, 
which also decays, any $c$ contribution to the 
Universe's energy density will rapidly dilute and the Universe will quickly enter a slow roll inflationary regime 
governed by the evolution of $\psi$. Nevertheless, kination should not be readily dismissed. 
Indeed, as we argue at the end of section~\ref{Results}, 
a pre-inflationary period of kination may produce interesting observable effects in the cosmic microwave background
on very large angular scales, at least if inflation does not last for too long.

%Add plots that show this transition and how fast it happens. 

%Add section explaining how we think c can be observed, how it modifies the power spectrum. 

\section{Results}
\label{Results}

In this section we show the results that a numerical analysis of Eqs.~(\ref{Friedmann.Equation.efolding.time}--\ref{Scalar.Equation.efolding.time}) yields. 
There are 4 free parameters that control the dynamics of this theory - 3 coupling constants $\alpha$, $\zeta$, $\lambda$ and the field-space angular momentum $c$. In this section we will see what role each of them plays in the predictions of the model.

We start with $\alpha$, as its role is simplest to understand. If we extract $\alpha$ out of all terms 
in the potential~(\ref{potential for psi}) we get,
\begin{equation} 
\label{potential}
{
V(\psi) = M_{\rm P}^4\left[\Big(\frac{9\xi^2}{4\alpha}+36\lambda\Big)\sinh^4(\psi)-\frac{3\xi}{4\alpha}\sinh^2(\psi)
                      +\frac{1}{16\alpha}\right]
\,,
}
\end{equation}
and looking at~\eqref{Scalar.Equation.efolding.time}, we see that -- for fixed $\xi$ and $\tilde\lambda\equiv\alpha \lambda $
--  the coupling $\alpha$ controls the size of the potential, 
{\it cf.} also~(\ref{V at origin}), and thus the Hubble parameter at the beginning
of inflation, as well as the scalar and tensor spectra of cosmological perturbations (recall that  
the corresponding amplitudes are, $\Delta_s^2\simeq H^2/(8\pi^2\epsilon M_{\rm P}^2)$ and 
$\Delta_t^2\simeq 2H^2/(\pi^2M_{\rm P}^2)\simeq 16\epsilon \Delta_s^2$,
respectively) and thus can be fixed by requiring the COBE normalization of the scalar power spectrum, 
$\ln(10^{10}\Delta^2_{s*})=3.089\pm0.036$.
% fixed at a comoving scale, $k_*=0.05~{\rm Mpc}^{-1}$.

\begin{figure} [ht]
	\centering
	\includegraphics[width=\linewidth]{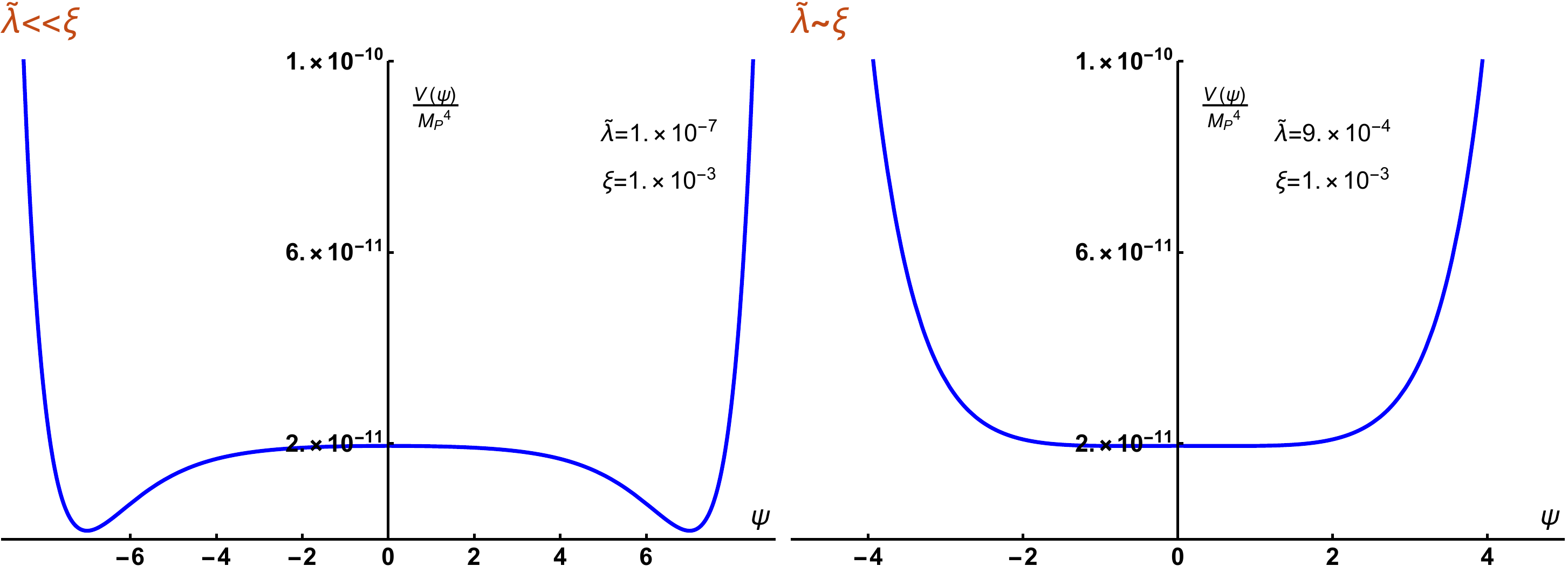}
	\caption{The hierarchy effect between $\tilde \lambda=\alpha \lambda$ 
and $\xi^2$. {\it Left panel.}
There are deep minima at $\psi=\pm\psi_m$, $\psi_m\gg 1$ when the hierarchy $1\gg\xi^2\gg \tilde\lambda$ is observed.
{\it Right panel.} When the hierarchy  $\xi^2\gg \tilde\lambda$ is not observed, 
the minima get close to the origin  and become very shallow, almost unobservable by eye, although they are still present.}
	\label{fig:potnomin}
\end{figure}

\begin{figure} [ht]
	\centering
	\includegraphics[width=0.9\textwidth]{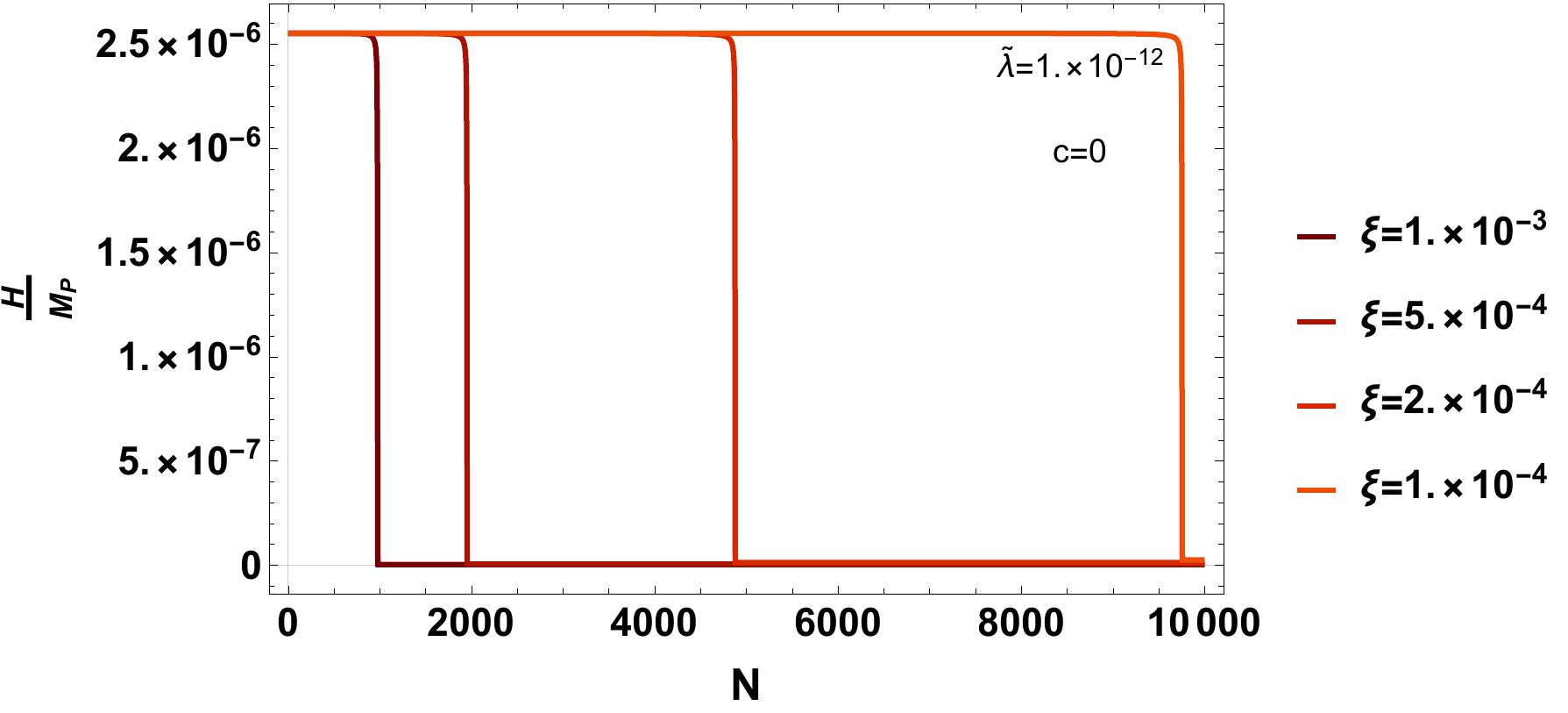}
	\caption{Increasing $\xi$ (for fixed $\alpha$ and $\lambda$) decreases the duration of inflation.}
	\label{HBetaDependence}
\end{figure}

The role of $\xi$ can be understood from the requirement that the potential should have 
a sufficiently large flat region around the origin. From Eqs.~(\ref{psi m}) and~(\ref{potential minimum}) 
we see that the minima will be at a large (super-Planckian) value if $\iota\gg1$.
In this case, inflation will be long (the total number of e-foldings $N_{\rm tot}$ 
will be much larger than the required 60) and these models are 
{\it large field inflationary models}. In the opposite limit when $\iota\ll 1$, $\psi_m\ll 1$, 
 inflation will typically be short ($N_{\rm tot}\ll 60$) and one gets 
a {\it small field inflation}. Obviously, viable inflationary models must satisfy $N_{\rm tot}\geq 60$ and belong to 
large field models, for which  $\iota\gg1$. As we will see below, it is also typically the case that $16\alpha\lambda\ll\xi^2$,
such that $\iota\simeq 1/\sqrt{6\xi}$, and~(\ref{psi m}) yields $\psi_m\simeq \ln(2\iota)\simeq   \frac{1}{2}\ln(2/(3\xi))$.  
In figure~\ref{fig:potnomin} we illustrate the two relevant cases, 
large field (left panel) and small field (right panel) inflationary potentials.
Upon solving the background equations of motion, 
one can confirm that increasing $\xi$ (for fixed $\lambda$ and $\alpha$),
decreases the duration of inflation, as can be seen in figure~\ref{HBetaDependence}. 

\bigskip

To summarize, the following picture has emerged: 
\begin{enumerate}

\item[1)]
The parameter $\lambda\ll 1$ controls the vacuum energy density at the end of inflation. 
More precisely, from Eq.~(\ref{potential minimum}) it follows that $\lambda/(\xi^2+16\alpha\lambda)=V_m/M_{\rm P}^4$ 
controls the post-inflationary vacuum energy density, 
which in the limit when $\xi^2\gg 16\alpha\lambda$ reduces to $\lambda/\xi^2$. 
This vacuum energy should be of the order of the electroweak energy density, 
$\rho_{\rm EW}\sim 10^{-66}~M_{\rm P}^4$, 
 implying that $\lambda$ ought to be extremely small, {\it i.e.} $\lambda\sim 10^{-66}\xi^2$.

\item[2)]
The parameter $\alpha\gg 1$ controls the amplitude of the scalar and tensor power spectra, 
as it fixes the value of $H$ at the beginning of inflation.
It is therefore fixed by the COBE normalization of the amplitude of scalar cosmological perturbations to be about,
$\alpha\sim 10^9$, where to get the estimate we took, $r=16\epsilon\simeq 0.003$ of the Starobinsky model.

\item[3)] 
The parameter $\xi\ll 1$ controls the duration of inflation, such that a small $\xi$ implies a long inflation (large field model); 
a large $\xi$ implies short inflation (small field model), see figure~\ref{fig:Hvolut}. More precisely, it is actually
$\iota^{-2}=6(\xi^2+16\alpha\lambda)/\xi$ that controls the duration of inflation, 
which in the limit when $\xi^2\gg 16\alpha\lambda$ reduces to $6\xi$. 

\end{enumerate}

Finally, one can argue that $16\alpha\lambda\ll\xi^2$ as follows. If this condition were not met,
would imply (from~(\ref{potential minimum})) a vacuum energy density at the end of inflation, 
$V_m\simeq (M_{\rm P}^4/16\alpha)\left[1-\xi^2/(16\alpha\lambda)\right]$, which is comparable to 
the vacuum energy at the beginning of inflation given by~(\ref{V at origin}).  
If this energy is to be almost compensated by the vacuum energy from 
the electroweak symmetry breaking, 
this would mean that inflation would have to happen at the electroweak scale. 
Inflation at the electroweak scale is possible, but it is much more fine tuned than inflation 
close to the grand unified scale, at which  
$H\sim 10^{13}~{\rm GeV}$, and thus theoretically disfavored.

\subsection{Model predictions for $n_s$ and $r$}
\label{Model predictions for ns and r}

In what follows, we present inflationary predictions of our model in slow roll approximation~\footnote{By this we mean 
that the formulas for $n_s$, $\alpha_s$ and $r$ that we use are the leading order in slow roll. The background evolution is solved however exactly.}. 
There are essentially four observables~\footnote{If one includes 
the running of the scalar and tensor spectral index then there are six observables.}
 from the Gaussian cosmological perturbations - from which two have been observed 
and for the other two there are limits and there are only limits on non-Gaussianities. 
The scalar and tensor spectrum can be written as, 
\begin{equation}
 \Delta_s^2(k) = \Delta_{s* }^2\left(k/k_*\right)^{n_s-1}
\,,\qquad 
 \Delta_t^2(k) = \Delta_{t* }^2\left(k/k_*\right)^{n_t}
\label{spetra}
\end{equation}
where $\Delta_{s* }^2$ is the amplitude of the scalar spectrum, which is fixed by the COBE normalization,
\begin{equation}
A_s\equiv  \Delta_{s*}^2 = (2.105 \pm 0.030)\times 10^{-9}
\,
\label{ns}
\end{equation}
and $n_s$ is the spectral slope defined such that, when $n_s=1$, the scalar spectrum is scale invariant.
Planck (and other available) data~\cite{Aghanim:2018eyx}  constrain $n_s$ at $k_*=0.05~{\rm Mpc}^{-1}$ as, 
\begin{equation}
 n_s = 0.9665\pm 0.0038 \quad (68\% \,{\rm CL})
\,.
\label{ns}
\end{equation}
In slow roll approximation, $n_s$ can be expressed in terms of slow roll parameters $\epsilon_1$ and 
$\epsilon_2 = (d/dN)\ln(\epsilon_1)$ as, 
\begin{equation}
 n_s  =1 -2\epsilon_1-\epsilon_2
\,,
\label{ns}
\end{equation}
where higher order (quadratic, {\it etc.}) corrections in slow roll parameters have been neglected.  
On the other hand, we have no measurements of the tensor spectrum. The current upper bound is 
expressed in terms of the scalar-to-tensor ratio, defined as, 
\begin{equation}
 r = \frac{\Delta_s^2}{\Delta_t^2} 
\,,
\label{r:rdef}
\end{equation}
which to leading order in slow roll parameters reads, $r=16\epsilon_1$, and it is constrained by the data 
at $k_*=0.002~{\rm Mpc}^{-1}$ as,
\begin{equation}
 r <  0.065 \quad (95\% \,{\rm CL})
\label{r: from data}
\end{equation}
Because $r$ has not yet been measured, there is no meaningful constraint on the tensor spectral index, $n_t=-2\epsilon_1=-r/8$.
Sometimes one also quotes limits on the running of the scalar spectral index, defined as 
$\alpha_s=dn_s/d\ln(k)$, which makes up the fifth Gaussian observable.
The current constraints on $\alpha_s$ are quite modest, $-0.013<\alpha_s<0.002$, and they are 
not strong enough to significantly constrain our model, in which $\alpha_s$ is second order in slow roll parameters 
and thus quite small. 

\bigskip

In what follows, we investigate how the model predictions depend on the parameters of the model $\xi$, $\lambda$ and $\alpha$. 
The dependence on $\alpha$ is the simplest, as it is completely fixed by the COBE normalization and by 
the scalar-to-tensor ratio $r$ as,
\begin{equation}
 \alpha =\frac{1}{24\pi^2 r A_{s*}}
\,.
\label{observational limit on alpha}
\end{equation}
Now, as we shall see below, $r$ can be well
 approximated by, $r\approx 0.003$ (with $\sim 30\%$  accuracy), implying that 
$\alpha\approx 7\times 10^8$.   

\begin{figure} [ht]
	\centering	
	\begin{subfigure}[b]{0.49\textwidth}
		\centering
		\includegraphics[width=\textwidth]{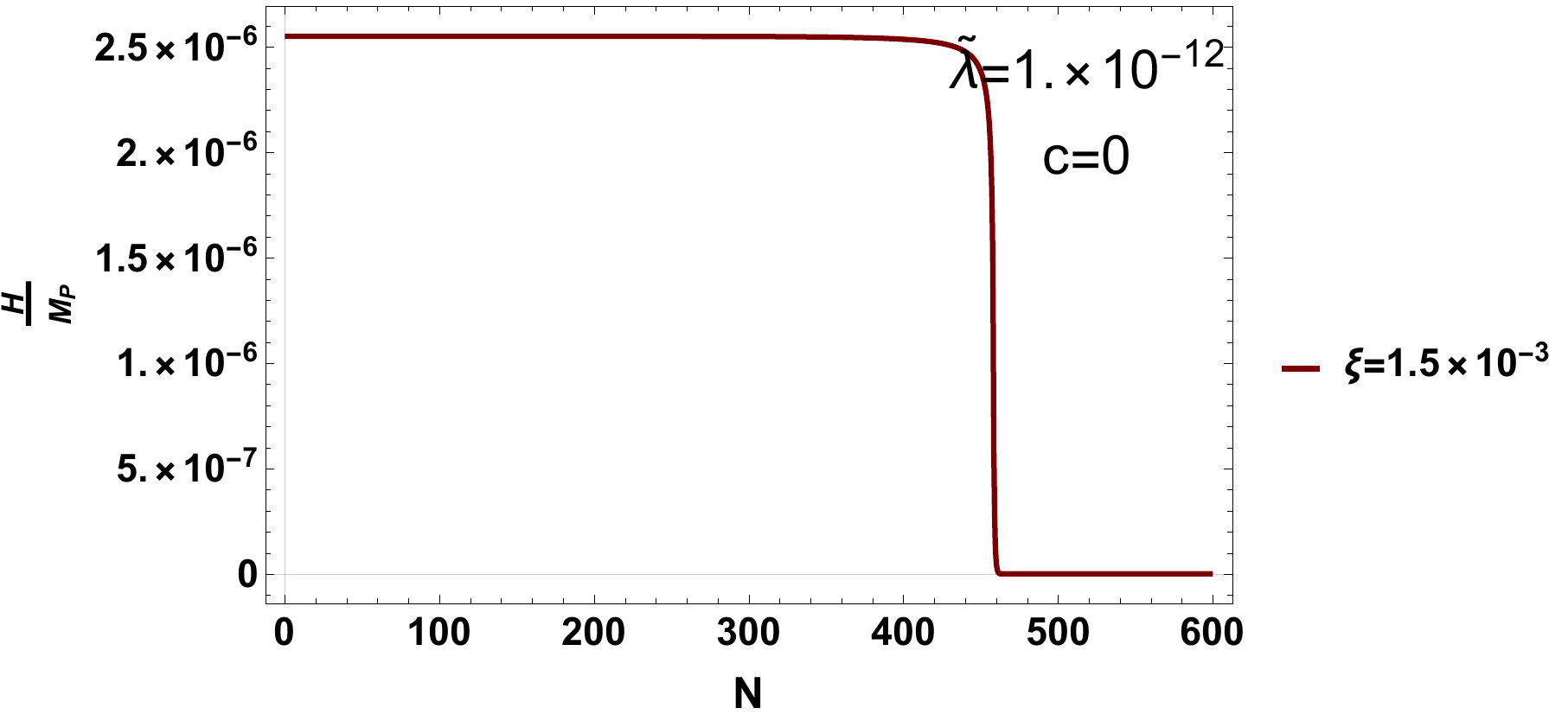}
		\caption{For $\xi = 1.5\cdot 10^{-3}$ inflation lasts for long time, much more than the required $60$ e-foldings.}
		\label{fig:Hvolution}
	\end{subfigure} \hfill
	\begin{subfigure}[b]{0.49\textwidth}
		\centering
		\includegraphics[width=\textwidth]{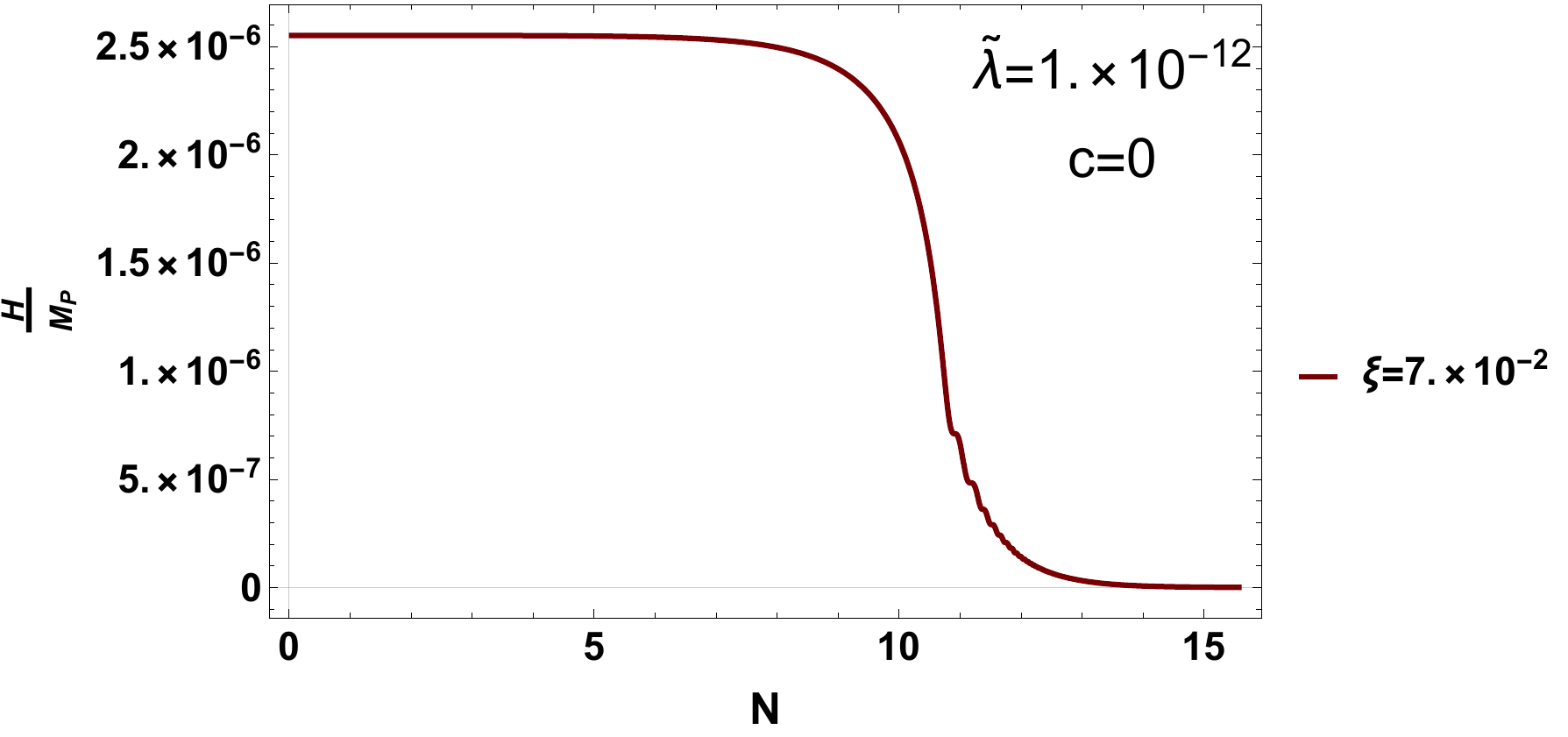}
		\caption{For a largish $\xi$ ($\xi = 0.07$) inflation does not any more last long enough as $N_{\rm tot}\simeq 10\ll 60$.}
		\label{fig:HvolutionBad}
	\end{subfigure}
	\caption{The duration of inflation in a large (left) and small (right) field inflationary model.}
	\label{fig:Hvolut}
\end{figure}
Next we look at the dependence on $\xi$. In figure~\ref{fig:Hvolut} we illustrate how the duration of inflation 
depends on $\xi$. The case $\xi\ll 1$ falls into the class of {\it large field models} (left panel) and one gets $N_{\rm tot}\gg 60$.
On the other hand, the case $\xi\lesssim 1$ is a {\it small field model} 
and yields only a few e-folds, {\it i.e.} $N_{\rm tot}\ll 60$ (right panel), making it not viable for inflationary model building.
We note that one gets enough e-foldings of inflation  ($N_{\rm tot}\sim 60$)  when $\psi_m\sim 3-4$, which belongs to the class of 
{\it intermediate field models}, in which during inflation the inflaton $\psi$ rolls over approximately one (unreduced) 
Planck mass, $\Delta\psi\approx \psi_m\sim m_{\rm P}=\sqrt{8\pi}M_{\rm P}$.

\begin{figure} [ht]
	\centering
	\includegraphics[width=10cm]{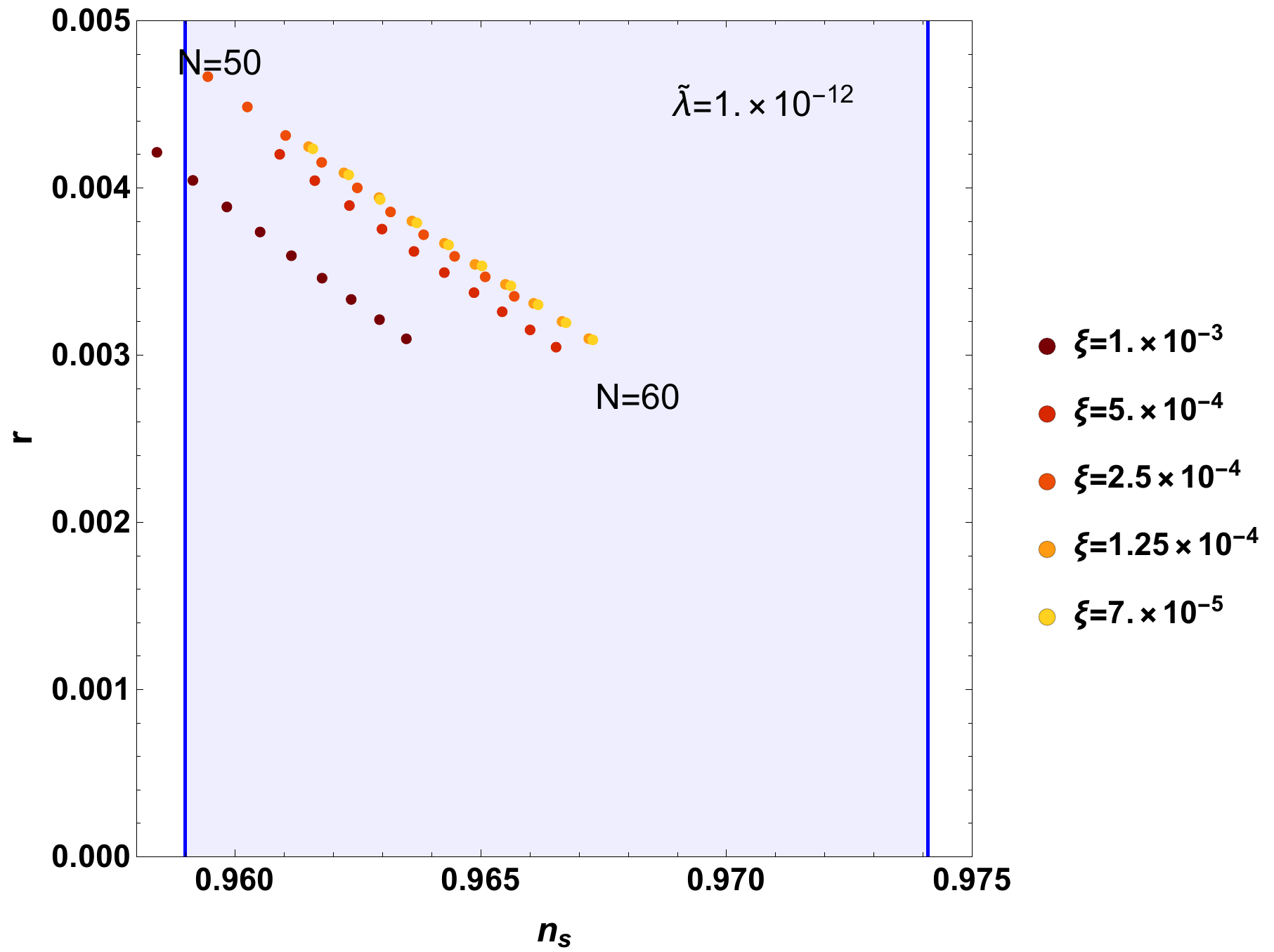}
	\caption{$r$ {\it vs} $n_s$ for varying $\xi$.  As  $\xi$ decreases,  $r$ and  $n_s$ approach those
of Starobinsky's inflation. As $\xi$ increases, $n_s$ and $r$ mildly decrease.}
	\label{BetaMultiplot}
\end{figure}
\begin{figure} [ht]
	\centering
	\includegraphics[width=10cm]{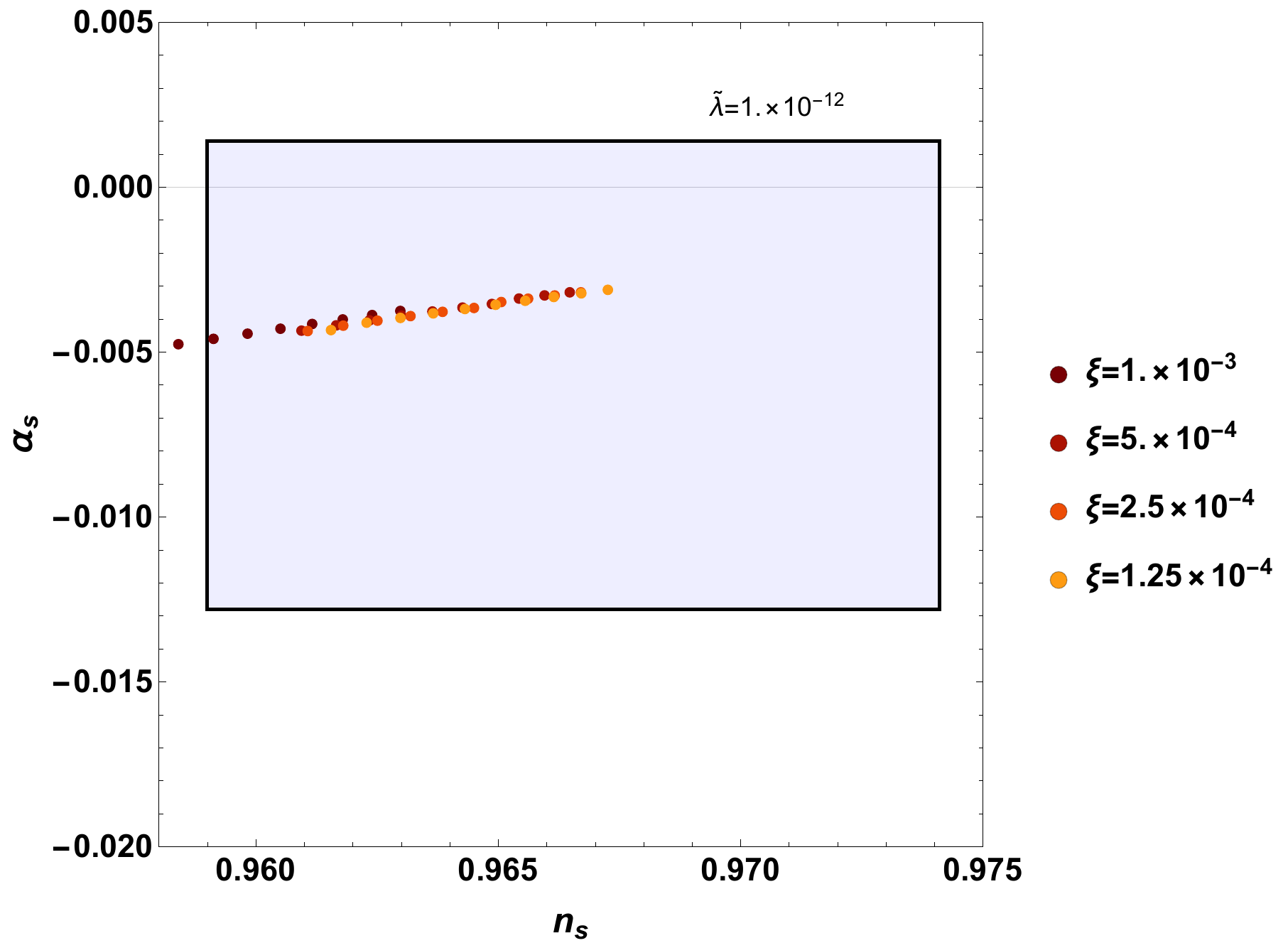}
	\caption{$\alpha_s$ {\it vs} $n_s$ for varying $\xi$.  As  $\xi$ decreases,  $\alpha_s$ does not vary much, while $n_s$ approaches the value of Starobinsky's inflation. 
	The rectangle limits this time denote the $1\sigma$ contour as given in~\cite{Aghanim:2018eyx}.}
	\label{BetaMultiplotAlpha}
\end{figure}

In figure~\ref{BetaMultiplot} we show how $n_s$ and $r$ change as $\xi$ varies for a fixed $\tilde\lambda\equiv \alpha\lambda$.
The figure shows that, in the limit when $\xi$ is very small, one reproduces the $n_s$ and $r$ of Starobinsky's model; increasing 
$\xi$ introduces a deviation from Starobinsky's model such that both $n_s$ and $r$ decrease. 
Note that  $\xi$ cannot be too large, since $n_s$ decreases as $\xi$ increases, eventually dropping outside the region of validity
of figure~\ref{BetaMultiplot}. 
In figure~\ref{BetaMultiplotAlpha}, we plot the running of the scalar spectral index, $\alpha_s$,
and the corresponding observational $2\sigma$ contours. As one can see, our model predicts a rather small $\alpha_s$, 
which is well within the experimental limits when $\xi\ll 1$.
\begin{figure} [ht]
	\centering
	\includegraphics[width=10cm]{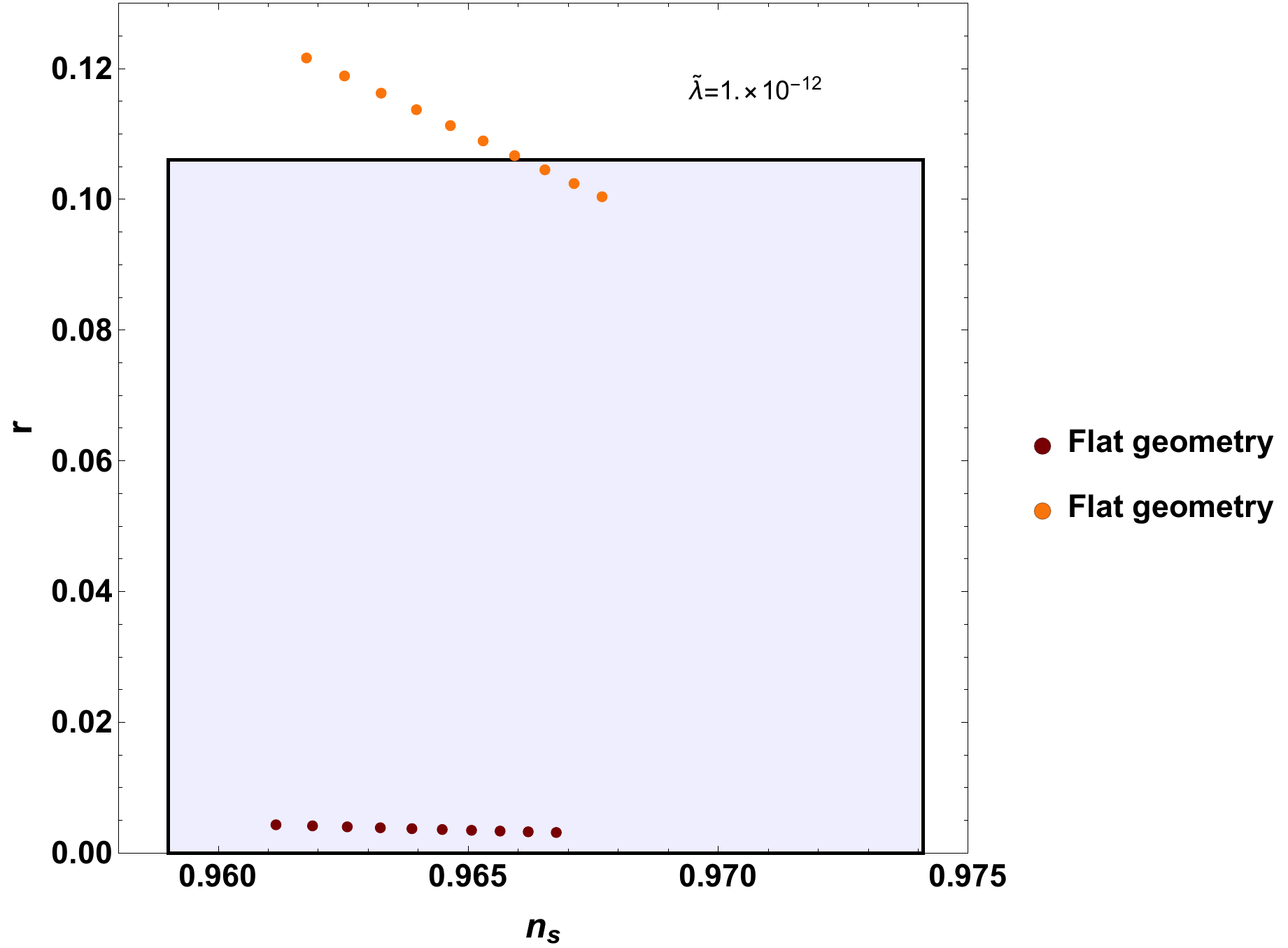}
	\caption{$r$ {\it vs} $n_s$ for $\xi = 0.25\times 10^{-3}$ in the two cases of hyperbolic and flat field space
	geometry. Here the box indicates the $2\sigma$ contour as given in~\cite{Aghanim:2018eyx}.
	As we can see the flat geometry exceeds the $95\%$ confidence level contours for $r$ for this particular value of $\xi$, 
	while the hyperbolic geometry sits comfortably at a low value of $r$. }
	\label{fieldgeometryeffect}
\end{figure}

In figure~\ref{fieldgeometryeffect} we show the effect of the hyperbolic field space
geometry on the tensor to scalar ratio. Compared to the case of flat geometry, the 
slow roll parameters stay smaller for a longer time~(in this sense the hyperbolic potential is 
more ``flat''), which eventually translates into a suppression in the tensor to scalar ratio.
For small values of $\xi\simeq 10^{-4}$ the flat space model is excluded, while the 
hyperbolic space model gives reasonable predictions as long as the condition $\xi^2\gg16\alpha\lambda$,
as described in section~\ref{Inflationary dynamics}, is satisfied.
%

%\begin{figure} [ht]
%	\centering
%	\includegraphics[width=7.5cm]{LambdaMultiplot.pdf}
%	\caption{The parameters $n_s$ and $r$ depend very weakly on $\lambda$, unless $\tilde\lambda\gtrsim \xi^2/16$.}
%	\label{LambdaMultiplot}
%\end{figure}
%
Finally, let us discuss the dependence on $\lambda$. As explained above, $\lambda$ primarily controls 
the vacuum energy at the end of inflation, and for that reason it ought to be small enough. 
When $\alpha\lambda \ll \xi^2/16$ the inflationary observables depend only very weakly on $\lambda$, and 
the dependence on $\lambda$ starts becoming significant as $\alpha\lambda \sim \xi^2/16$ or larger.
However, when the latter condition is satisfied, the vacuum energy left when $\psi$ settles on its minimum
is big enough to quickly dominate the universe and drive a second phase of accelerated expansion. 
In this case, the model would predict eternal inflation and would be as such ruled out. Hence we must require $\lambda\ll\xi^2/16$, in which limit any dependence on $\lambda$ essentially drops out.

\bigskip

\subsection{The role of the flat direction} 
\label{The role of the flat direction chi} 

\begin{figure} [ht] 
	\centering	
		\includegraphics[width=0.6\textwidth]{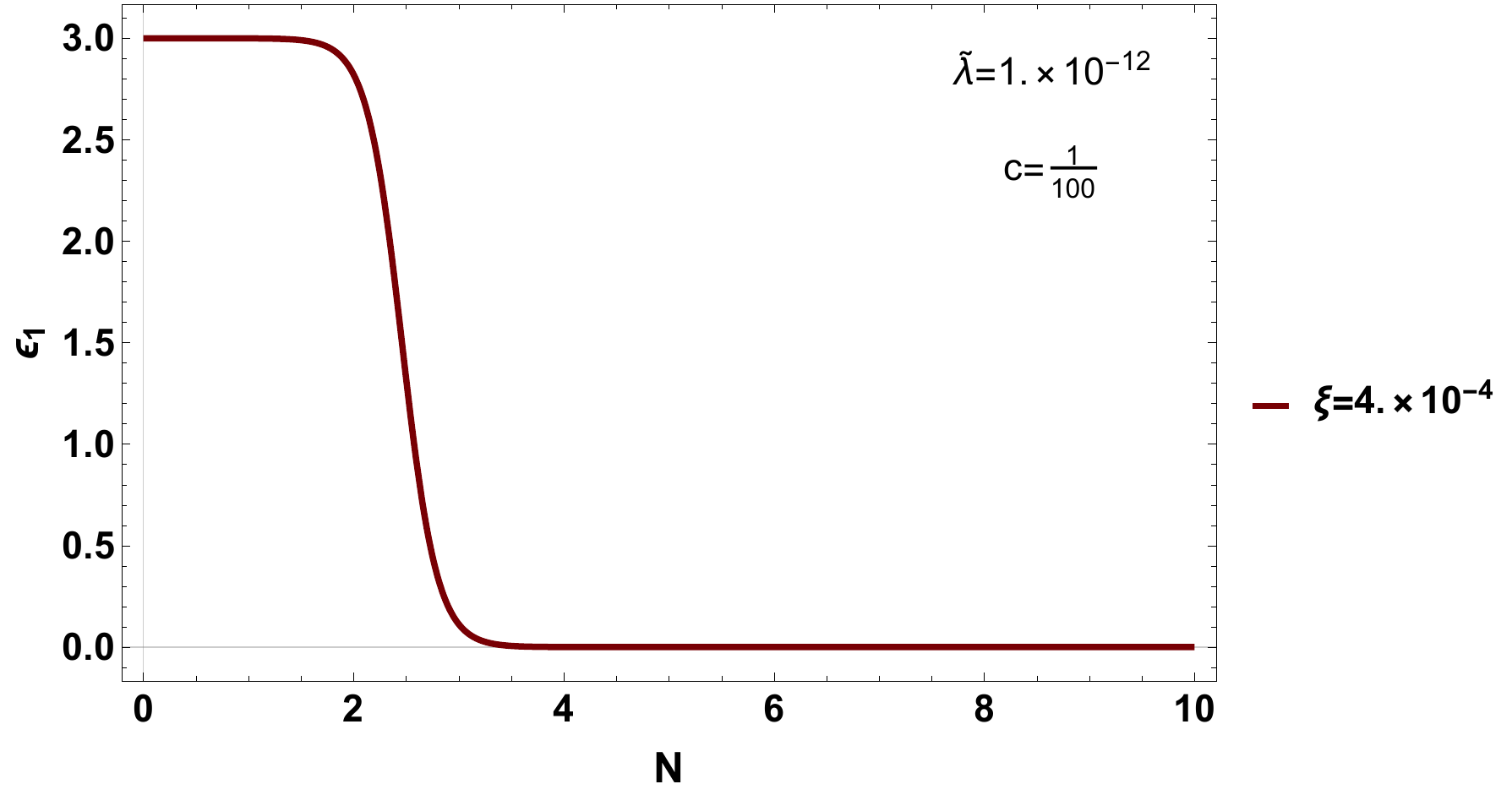}
	\caption{The parameter $c$ introduces a brief period of kination, after which the universe transits to a quasi de Sitter epoch.}
	\label{fig:Kination}
\end{figure}

So far we have not yet discussed how the model predictions depend on the dynamics of the flat direction $\chi$. 
Unless the initial kinetic energy stored in $\chi$ is large, it will not affect the above analysis in any significant manner.
Consider however the case when the initial kinetic energy in $\chi$ is large. This is equivalent to 
taking the parameter $c$ defined in Eq.~(\ref{c: definition}) to be initially large. One can easily convince oneself 
that in this case the energy density of the Universe will  early on  scale as, $\propto c^2/[a^6\cosh^2(\psi)]$, 
which corresponds approximately to a period of kination, during which kinetic energy of a scalar field dominates
and $\epsilon_1\approx 3$. A brief period of kination 
followed by slow roll inflation is clearly visible in figure~\ref{fig:Kination}.
From the value of the Hubble parameter at the beginning of inflation, $H_I\simeq 10^{13}~{\rm GeV}$, we easily get 
an estimate for the maximum number of e-foldings in kination, 
\begin{equation}
 (N_{\rm kin})_{\rm max}\lesssim \frac16\ln\left(\frac{M_{\rm P}^2}{H_I^2}\right)\simeq 4
\,,
\label{maximum kination}
\end{equation}
where the estimate is obtained by assuming that the initial energy density in kination is at most Planckian.

\begin{figure} [ht]
	\centering
	\includegraphics[width=0.7\linewidth]{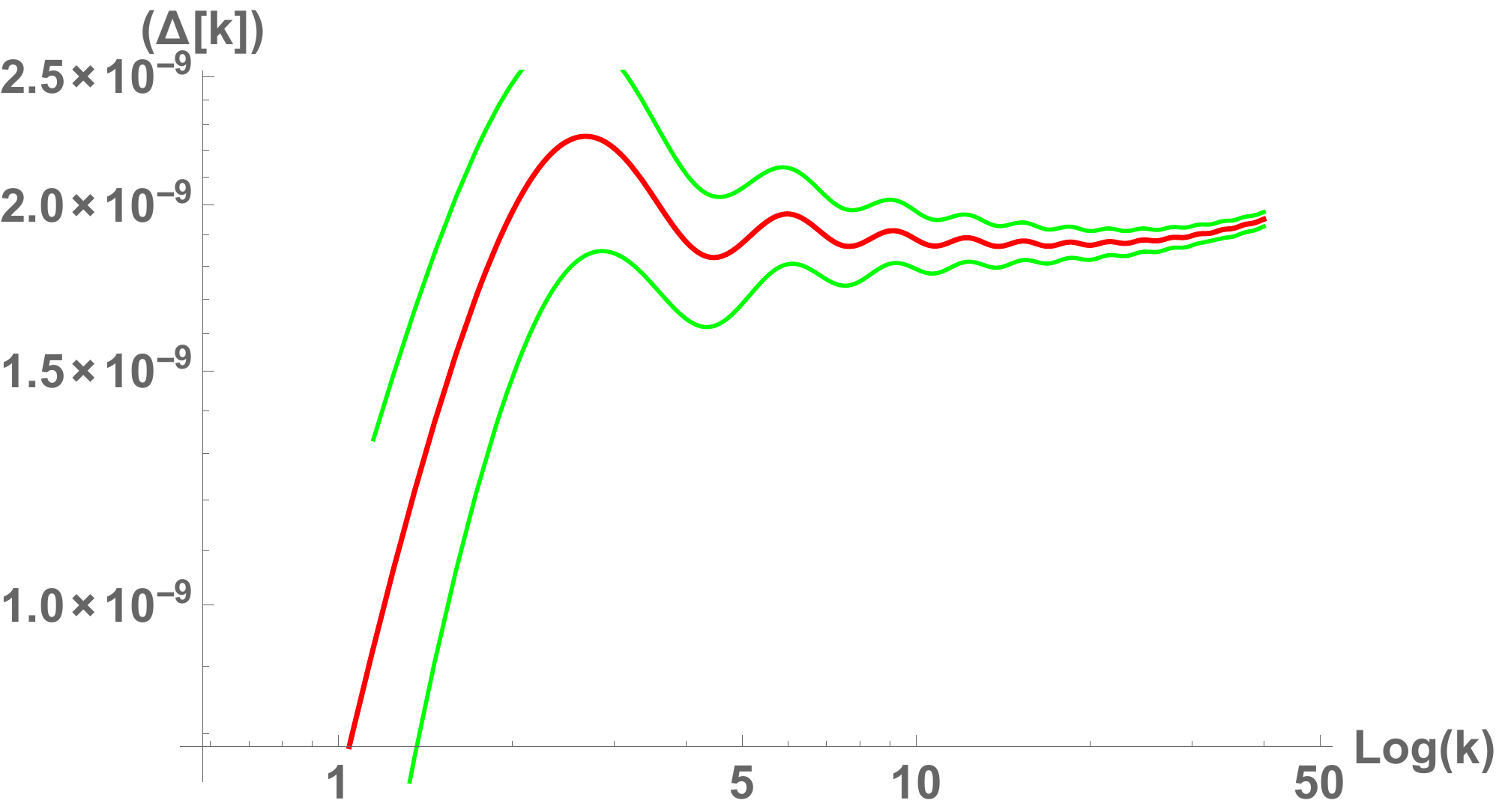}
	\caption{Power Spectrum in the sudden matching approximation. Notice the sudden 
drop in power on the largest scales. The upper and lower green lines represent the 
	cosmic variance contours.}
	\label{fig:pspectrum}
\end{figure}

A proper analysis of the spectrum of scalar cosmological perturbations requires solving for small perturbations 
of two fields, where the adiabatic mode is a linear combination of the two fields, fluctuations of $\chi$ and of $\psi$.
During kination mostly $\chi$ will source scalar cosmological perturbations, during inflation it will be $\psi$, while 
at the transition from kination to inflation it will be a linear combination of the two. 
Therefore as a rough approximation one can assume: the fluctuations of the $\chi$ field source the adiabatic mode during kination,
 the fluctuations of the $\psi$ field source it during inflation, and the transition period is instantaneous.~\footnote{
This approximation neglects the effect of turning of the trajectory in field space~\cite{Achucarro:2010jv}, 
and a proper understanding of this effect we postpone for a future publication.
}
An inspection of figure~\ref{fig:Kination} suggests that the transition lasts for about one e-folding, of equivalently 
about one Hubble time, $\Delta t_{\rm transition} \sim 1/H$.
This then  means that sub-Hubble modes ($k/a\gg H$) behave adiabatically, 
{\it i.e.} their matching leads to an exponentially suppressed mixing 
between positive and negative frequency modes, and thus to an exponentially suppressed particle production and 
the amplification for these modes can be neglected. 
On other other hand, super-Hubble modes ($k/a\ll H$) can be treated in the sudden matching approximation, 
such that these modes inherit the highly blue spectrum from kination, $\Delta_s\propto (k/k_*)^3$, for which $n_s\simeq 4$.
The resulting power spectrum is shown in figure~\ref{fig:pspectrum}.

As it was originally pointed out by Starobinsky~\cite{Starobinsky:1992ts} (see also~\cite{Brooker:2015iya}), 
apart from a {\it break} in its slope, the scalar power spectrum 
also exhibits a {\it memory} effect, manifested as {\it damped oscillations} which propagate into quite large momenta. 
These oscillations can be clearly seen in figure~\ref{fig:pspectrum}, 
where for definiteness we have assumed that the Hubble parameter at the matching 
equals to $k\sim 3 H_0$ (which roughly corresponds to the CMB multipole $l\sim 3$). The scale at which 
the spectrum breaks can be shifted left or right by a suitable change in the Hubble parameter at the matching, or equivalently 
by changing the duration of the inflationary phase.
Even though 
the oscillations are generated by the matching at $k\sim 3 H_0$, 
they are visible all the way to $k\sim 30H_0$, or equivalently $l\sim 30$.
In order to better understand whether these oscillations are strong enough to be detectable we show the upper and lower 
contours obtained by adding and subtracting the cosmic variance  (shown as green lines).
We emphasize that, even though the size of the oscillations is smaller than the cosmic variance, their cumulative effect might be 
statistically significant and one should look for their effect in the data.  

As it is known from the Planck Collaboration~\cite{Aghanim:2018eyx} 
(see also the earlier observations of COBE and WMAP~\cite{Bennett:1996ce,Spergel:2003cb}), 
the temperature anisotropy spectrum of CMB has a dip in the low $l$ part of the spectrum which has statistical significance of 
$2-3 \sigma$ which could be explained by a pre-inflationary period of kination. Such an explanation is viable only if inflation 
lasts for about 60 e-foldings. While this may seem like a tuning of inflation, we point out that 60 e-foldings are obtained 
in our model if the field rolls during inflation by about 1 (unreduced) Planck scale, which might be the natural scale of inflation.
Namely, large field inflationary models might be hampered by quantum operators of large dimensions.  In our model however,
such operators may be forbidden by conformal symmetry; for a recent discussion of the role of such operators for inflation 
see~\cite{Marunovic:2016reh}.

\section{Quantum corrections at 1-loop }
\label{Quantum.Corrections}

In this section we investigate the quantum corrections to the 
action~(\ref{Weyl.invariant.action}) and show that the model, 
and in particular the hierarchy required for the inflationary 
observables to match the Planck data, {\it i.e.} $\lambda\ll\xi\ll \alpha$,
is preserved by quantum corrections. 

So far we have learned that the theory~(\ref{einstein.frame.action}) contains one massive scalar field $\phi$ 
(defined in Eq.~(\ref{constraint.1})), $\mathcal{N}-1$ massless Goldstone bosons, one flat direction (denoted by $\chi$), 
one conformal gauge degree of freedom (denotes by $\omega$),
one vector degree of freedom (which can be removed by setting $\sigma=0$) and one massless, 
spin two degree of freedom (the graviton). 
For simplicity here we shall consider the O(1) case, in which there are no Golstone bosons.
Furthermore, the contribution of the massless scalar $\chi$ to the (infrared) effective action 
is suppressed when compared to that of the massive scalar and therefore can be neglected. 

Let us now consider the one loop contribution of the massive scalar $\phi$.
As it is well known, the quantum effective action that 
describes the quantum theory is given, at one loop,
by the classical action plus the term $(i/2)\text{Tr}\log (i\Delta)^{-1}$,
where $i\Delta$ is the scalar propagator of the theory, which for the 
action~(\ref{Weyl.invariant.action}) is given by the solution of,
\begin{equation}
\label{propagator.equation}
\left [ {\nabla}{}_\mu{\nabla}{}^\mu + \xi R-3\lambda \phi^2\right ]i\Delta(x;y)
        = i\frac{\delta^D(x\!-\!y)}{\sqrt{-g(x)}}\,,
\end{equation}
where $\phi$ is the value of the~(possibly space-time dependent)
scalar field condensate, {\it i.e.} $\phi=\langle\hat\phi\rangle$ and $D$ is the dimension of space-time.

While the effective action is highly non-local~\cite{Barvinsky:1985an}, its infrared limit 
(in which $\xi R,\lambda \phi^2\gg \|\Box\|$) is quite simple. Indeed, 
the unregularized effective action is given by,
\begin{eqnarray}
\Gamma\!&=&\!  \int \frac{\text{d}^D x}{(4\pi  )^{D/2}} \sqrt{-g}\Bigg[ \left (-\left(\xi - \frac{1}{6}\right ){R}+\lambda \phi^2\right )^{\frac{D}{2}} \Gamma\left (-\frac{D}{2}\right ) \\
\!&-&\! \frac{1}{3} \left (\frac{1}{180} {R}{}_{\alpha(\beta\gamma)\delta}{R}{}^{\alpha(\beta\gamma)\delta}-\frac{1}{180} {R}{}_{(\alpha\beta)}{R}{}^{(\alpha\beta)} +\frac{(D-2)^2}{48} {\cal T}_{\alpha\beta}{\cal T}^{\alpha\beta}\right )\\
\!&\times&\! \left (-\left(\xi - \frac{1}{6}\right ){R}+\lambda \phi^2\right )^{\frac{D-4}{2}}  \Gamma\left (2-\frac{D}{2}\right ) \Bigg ]\,.
\end{eqnarray}
From this action we can read off the quantum corrections to the interaction
vertices. We find,
\begin{eqnarray}
\label{quantum.corrections.couplings} 
\delta\alpha \!&=&\! \frac{\left(\xi-\frac{1}{6}\right)^2}{16\pi^2}
  \log\left[\frac{-\left(\xi - \frac{1}{6}\right ){R}+3\lambda \phi^2}{\mu^2}\right]
\,,\\
\delta\xi \!&=&\! -\frac{\left (\xi-\frac{1}{6}\right ) \lambda}{8\pi^2}
 \log\left[\frac{-\left(\xi - \frac{1}{6}\right ){R}+3\lambda \phi^2}{\mu^2}\right]
\,,\\
\delta\lambda \!&=&\! -\frac{\lambda^2}{16\pi^2} \log\left[\frac{-\left(\xi - \frac{1}{6}\right ){R}+3\lambda \phi^2}{\mu^2}\right]
\,,
\end{eqnarray}
where we assumed that, $-\left(\xi - \frac{1}{6}\right ){R}+3\lambda \phi^2>0$.
This shows that the quantum corrections at one loop can maintain 
a specific hierarchy between the coupling constants, 
namely,
\begin{equation}
\label{Hierarchy} 
  \lambda \ll \xi \ll \alpha\,, \implies \delta\lambda \ll \delta\xi\ll \delta \alpha\,,
\end{equation}
which also happens to be the required hierarchy to obtain 
convincing inflationary predictions, as we showed in the previous 
section. 

This discussion, however, does not take into account possible 
quantum gravitational corrections. While to reliably estimate their 
contribution one should set up a perturbative quantum gravity calculation,
such as the one that has performed in~\cite{Salvio:2017qkx},
we also expect these contribution to yield corrections that are suppressed 
at energy scales below the Planck energy. As long as the space-time curvature remains
within such a bound, therefore, we do not expect quantum gravitational correction 
to substantially change the conclusions of this section.

%CItation of the Jan Weenink paper still missing here, we discussed about it but i could not find the relevant paper. 

\section{Conclusions}
 
We investigate a simple inflationary model~(\ref{Weyl.invariant.action}) 
which exhibits local Weyl (or conformal) symmetry at the classical level which is realized by 
the space-time torsion. We show that in its simplest realization the model contains two scalar fields and one massless spin two
field. One of the scalars corresponds to a flat direction $\chi$, which is a remnant of conformal symmetry and therefore 
the flatness is protected from quantum corrections. The other scalar ($\psi$)  is massive and can support inflation. 
The noteworthy property of the model is a negative configuration space curvature\footnote{A similar feature 
flanks the so-called $\alpha$-attractor models~\cite{Kallosh:2013maa} constructed from a supergravity model.} 
which is responsible for the flattening 
of the effective potential for $\psi$ which is crucial for obtaining a long lasting inflation. 
While this feature can be present in other realisation of the spontaneous symmetry breaking pattern such 
as in the model of~\cite{Ferreira:2018qss}, it appears that the maximal symmetry is related to the enhanced, local,
Weyl symmetry. The predictions of low tensor to scalar ratio, which seems typical for Weyl invariant models, seems
to be in line with the results found in~\cite{Ozkan:2015kma}. 
Our analysis shows that, for a typical choice of parameters of the model, the model approximately reproduces 
the results of Starobinsky inflation, which again seem to appear as a `universal attractor'. 
However, variation of the coupling constants yields significant variation of 
in the predictions of the model as regards the scalar spectral index $n_s$ and the tensor-to-scalar ratio $r$, 
which can be used to test the model by the next generation of CMB satellite missions such as COrE~\cite{DiValentino:2016foa}.
Therefore, for quite a large range of couplings~(namely $\xi<0.01$, $\lambda\ll\xi$), our model is viable in light of the currently available data. 
We also point out that, if a lot of energy is initially stored in the flat direction, the universe will undergo a short period of kination,
followed by quasi-de Sitter inflation. Such a sequence is characterized by a lack of power in low momentum modes, 
a break in the power spectrum and a memory manifested as damped oscillations in the power spectrum. 

\section*{Appendix: A comparison between the breaking of Weyl symmetry and Abelian gauge symmetry}
\label{Higgs-Weyl-Comparison}

 It is important to understand that there are differences between the breaking of local Weyl symmetry and 
that of local Abelian gauge symmetry. The purpose of this Appendix is to underpin the similarities and differences between
the two. Our starting point is the Einstein frame action~(\ref{einstein.frame.action}) and the action for the 
Abelian-Higgs model (also known as scalar quantum electrodynamics, or short SQED),
\begin{equation}
S_{\rm SQED} \int \text{d}^4x \sqrt{-g} \Bigg[
  -\frac14 g^{\mu\nu}g^{\rho\sigma}F_{\mu\nu}F_{\rho\sigma} 
          - g^{\mu\nu}\left\{(\partial_\mu-ieA_\mu)\varphi^*(\partial_\nu+ieA_\nu)\varphi\right\}
           -V(\phi)
\Bigg ]\,,
\label{action: Abelian Higgs}
\end{equation}
where $e$ is the gauge coupling (electric charge), 
$F_{\mu\nu}=\partial_\mu A_\nu - \partial_\nu A_\mu$ is the field strength associated to 
the Abelian gauge field $A_\mu$, $\varphi$ is a complex scalar and 
\begin{equation}
 V(\varphi) = -\mu^2\varphi\varphi^* + \lambda(\varphi\varphi^*)^2
\label{Abelian Higgs potential}
\end{equation}
is the potential, which for $\mu^2>0$ exhibits spontaneous symmetry breaking that `breaks' the local gauge symmetry. 

Note first that the action~(\ref{einstein.frame.action}) is invariant ({\it i.e.} it transforms into itself) 
under local Weyl transformations, 
\begin{equation}
\label{field.redefinitions}
 \omega\;\rightarrow \; \omega e^{-\zeta(x)}
\,,\phi^I \;\rightarrow \;e^{-\zeta(x)}\hat\phi{}^I
\,, {\cal T}_\mu  \;\rightarrow \;{\cal T}_\mu + \partial_\mu \zeta(x)
\,,
g_{\mu\nu}  \;\rightarrow \; e^{2\zeta(x)} g_{\mu\nu}\,,
\end{equation}
where $\zeta(x)$ is an arbitrary (regular) function of space and time. This means that a suitable choice of $\zeta$ can 
fix $\omega$ to a nonvanishing constant, which defines the Planck scale $M_{\rm P}$. This completely fixes 
 Weyl symmetry. 

 Analogously, the Abelian-Higgs action~(\ref{action: Abelian Higgs}) is invariant under local gauge transformations, 
\begin{equation}
  A_\mu \;\rightarrow \;A_\mu + \partial_\mu \tilde\zeta
\,,\qquad \varphi \;\rightarrow\; {\rm exp}[-{ie\tilde\zeta(x)}]\varphi
\,,
\label{gauge transformations}
\end{equation}
where $\tilde\zeta=\tilde\zeta(x)$ is an arbitrary scalar gauge function.

The equation of motion for the torsion trace is obtained by varying the action~(\ref{einstein.frame.action}), 
\begin{equation}
\label{Torsion.Trace.EOM} 
\overset{\circ}{\nabla}_\mu \overset{\circ}{\nabla}{}^\mu{\cal T}_\nu  
     - \overset{\circ}{\nabla}{}^\mu\overset{\circ}{\nabla}{}_\nu {\cal T}_\mu 
  -(6\omega^2 + \phi^2) {\cal T}_\nu = \frac{1}{2} \partial_\nu (\phi^2+6\omega^2)\equiv {\cal J}_\nu
\,, \quad 
{\cal J}_{[\nu,\mu]}= 0
\,.
\end{equation}
Notice that the source current ${\cal J}_\nu$ is purely longitudinal, 
which is opposite to what happens in gauge theories. 
Indeed, the equation of motion of the gauge field implied by the action~(\ref{action: Abelian Higgs}) reads,  
\begin{equation}
\label{gauge.field.EOM} 
\overset{\circ}{\nabla}{}^\mu\overset{\circ}{\nabla}_\mu A_\nu 
 -   \overset{\circ}{\nabla}{}^\mu\overset{\circ}{\nabla}{}_\nu A_\mu - 2e^2\varphi\varphi^* A_\nu 
 = ie\left[\varphi\partial_\nu \varphi^*-\varphi^*\partial_\nu \varphi\right]
\equiv J_\nu
\,,\quad 
\overset{\circ}{\nabla}{}^\mu J_\mu =0 \,,
\end{equation}
where the scalar electromagnetic current $J_\mu$ is purely transverse. (The current transversality condition 
is a consistency condition that can be traced back to the gauge symmetry: since a massive gauge field contains 
at most 3 physical degrees of freedom, the current $J_\mu$ can have at most 3 independent components.)
 
 The first difference to notice in Eq.~(\ref{Torsion.Trace.EOM}) and Eq.~(\ref{gauge.field.EOM})
is that, in~(\ref{Torsion.Trace.EOM}), the effective mass does not vanish at zero
scalar field condensate, $\phi\rightarrow 0$. 
This is due to the gauge fixing condition, $\omega^2 = 4\alpha R + \xi \phi^2 \rightarrow M^2_P$,
which guarantees that the curvature condensate does not vanish when $\phi$ vanishes.

In order to reduce it to the backbones, let us recast the gauge field equation~(\ref{gauge.field.EOM}) 
  in flat space (Minkowski) limit $g_{\mu\nu} = \eta_{\mu\nu}$.
It is instructive to study~(\ref{gauge.field.EOM}) in the flat space-time limit, 
and assume that the scalar condensate is constant, such that~(\ref{gauge.field.EOM})  reduces to a Proca theory 
with a mass term given by, 
\begin{equation}
 M^2_A \equiv 2e^2\langle\varphi\varphi^*\rangle
\,.
\label{Proca mass}
\end{equation}
Acting suitable derivative operators on the Proca equation~(\ref{gauge.field.EOM})
 separates it into transverse and longitudinal equations as follows, 
\begin{eqnarray}
\label{transverse.equation.proca}
 (\partial^2-M^2_A )\partial_{[\mu} A_{\nu]} &=& \partial_{[\mu} J_{\nu]}
\,,\quad 
J_{\nu}= ie\left\langle\varphi\partial_\nu \varphi^*-\varphi^*\partial_\nu \varphi\right\rangle
\\
\label{longitudinal.equation.proca} 
  M^2_A \partial^{\nu} A_\nu &=& 0
\,,\quad 
(\partial_\nu M_A^2=0)
\,,
\end{eqnarray}
which tell us that (if $\partial_\nu M_A^2=0$)
the three propagating degree of freedom are transverse,
in the Lorenz sense, and massive. 
Notice that the Lorenz condition,  $\partial^{\nu} A_\nu= 0$, is exact as long as
 $M_A^2\neq 0$ and $\partial_\nu M_A^2=0$.

In the case of Weyl symmetry breaking, upon following an analogous procedure, one obtains 
\begin{eqnarray}
\label{transverse.equation.WeylSSB} 
(\partial^2-M^2_T)  \partial_{[\mu} {\cal T}_{\nu]} =2(\partial^2 {\cal T}_{[\mu}-\partial_{[\mu} \partial^{\lambda} {\cal T}_\lambda) {\cal T}_{\nu]}\,,\\
\label{longitudinal.equation.WeylSSB}
 -M^2_T  \partial^{\nu} {\cal T}_\nu 
         =\frac{1}{2} \partial^2 \langle\phi^2\rangle+\big(\partial^\nu \langle\phi^2\rangle\big) T_\nu
\,,\quad M^2_T = 6 M_{\rm P}^2 + \langle \phi^2\rangle
\,,
\end{eqnarray}
which tell us that the transverse modes are only sourced by higher order interactions
with the longitudinal mode~({\it i.e.} they are not sourced at linear order), while
the longitudinal degree of freedom is sourced at linear order. This is to be 
contrasted with the Abelian gauge theory, in which the transverse modes are sourced 
in the linearised theory, while the longitudinal mode is source-free at leading order.

This analysis gives a more formal justification of the {\it Ansatz} we used in section~\ref{geometric.theory.review}, 
${\cal T}_\mu = \partial_\mu \phi^0$, and justifies our proposition to take 
the longitudinal component of the torsion trace as an effective scalar degree of freedom. 
Indeed, Eq.~(\ref{longitudinal.equation.WeylSSB}) tells us that the scalar
$\phi^0$ mixes with the other scalar degree of freedom, and does not 
lead to a Lorenz condition as it is the case in the massive 
Proca theory. This scalar is the Goldstone mode of the broken local dilatation symmetry,
and mixes nonlinearly with all the scalars of the original theory according 
 to~(\ref{last.change}), which is the field redefinition that diagonalizes the field space metric.

While instructive, the above analysis is not rigorous. A rigorous analysis would entail
a proper (Dirac) analysis of constraints and dynamical degrees of freedom, and we leave it for future work.

\end{document}